\documentclass[fleqn,usenatbib]{mnras}

\usepackage{newtxtext,newtxmath}

\usepackage[T1]{fontenc}
\usepackage{ae,aecompl}

\usepackage{graphicx}	
\usepackage[utf8]{inputenc}
\usepackage[italian,english]{babel}
\usepackage{booktabs}
\usepackage{newtxtext,newtxmath}
\usepackage[dvipsnames]{xcolor}
\usepackage{subcaption}
\providecommand{\pgfsyspdfmark}[3]{}
\usepackage[english]{babel}
\usepackage{footmisc}

\defcitealias{Cannarozzo2021}{Cannarozzo (2021)}

\newcommand{\msun}{\mathrm{M_\odot}}
\newcommand{\kpc}{\mathrm{kpc}}
\newcommand{\kms}{\mathrm{km\,s^{-1}}}
\newcommand{\angstrom}{\text{\normalfont\AA}}

\newcommand{\logm}{\log M_*}

\newcommand{\reff}{R_{\mathrm e}}

\newcommand{\mstar}{M_*}

\newcommand{\dprime}{{\prime\prime}}
\newcommand{\mpc}{{\mathrm{Mpc}}}

\DeclareRobustCommand{\VAN}[3]{#2} 

\title[In-situ and Ex-situ Star Formation in ETGs]{The Contribution of In-situ and Ex-situ Star Formation in Early-Type Galaxies: MaNGA versus IllustrisTNG}

\author[C. Cannarozzo et al.]{\noindent
Carlo Cannarozzo$^{1,2,3}$\thanks{E-mail: ccannarozzo@astro.unam.mx},
Alexie Leauthaud$^{4}$,
Grecco A. Oyarzún$^{4}$,
Carlo Nipoti$^{2}$,\newauthor
Benedikt Diemer$^{5}$,
Song Huang$^{6}$,
Vicente Rodriguez-Gomez$^{7}$,
Alessandro Sonnenfeld$^{8}$,\newauthor
and Kevin Bundy$^{4,9}$
\\
$^{1}$Instituto de Astronomía, Universidad Nacional Autónoma de México, A. P. 70-264, 04510 CDMX, México\\
$^{2}$Dipartimento di Fisica e Astronomia "Augusto Righi", Alma Mater Studiorum Università di Bologna, via Piero Gobetti 93/2, I-40129 Bologna, Italy\\
$^{3}$INAF - Osservatorio di Astrofisica e Scienza dello Spazio di Bologna, Via Piero Gobetti 93/3, I-40129 Bologna, Italy\\
$^{4}$Department of Astronomy and Astrophysics, UCO/Lick Observatory, University of California, 1156 High Street, Santa Cruz, CA 95064, USA\\
$^{5}$Department of Astronomy, University of Maryland, College Park, MD 20742, USA\\
$^{6}$Department of Astrophysical Sciences, Princeton University, 4 Ivy Lane, Princeton, NJ 08544, USA\\
$^{7}$Instituto de Radioastronomía y Astrofísica, Universidad Nacional Autónoma de México, A. P. 72-3, 58089 Morelia, Mexico\\
$^{8}$Leiden Observatory, Leiden University, Niels Bohrweg 2, 2333 CA Leiden, The Netherlands\\
$^{9}$UCO/Lick Observatory, University of California, Santa Cruz, 1156 High Street, Santa Cruz, CA 95064, USA
}

\date{Accepted 2022 October 14. Received 2022 October 13; in original form 2022 May 02}

\pubyear{2022}

\begin{document}
\label{firstpage}
\pagerange{\pageref{firstpage}--\pageref{lastpage}}
\maketitle

\begin{abstract}
We compare stellar mass surface density, metallicity, age, and line-of-sight velocity dispersion profiles in massive ($M_*\geq10^{10.5}\,\msun$) present-day early-type galaxies (ETGs) from the MaNGA survey with simulated galaxies from the TNG100 simulation of the IllustrisTNG suite. We find an excellent agreement between the stellar mass surface density profiles of MaNGA and TNG100 ETGs, both in shape and normalisation. Moreover, TNG100 reproduces the shapes of the profiles of stellar metallicity and age, as well as the normalisation of velocity dispersion distributions of MaNGA ETGs. We generally also find good agreement when comparing the stellar profiles of central and satellite galaxies between MaNGA and TNG100. An exception is the velocity dispersion profiles of very massive ($M_*\gtrsim10^{11.5}\,\msun$) central galaxies, which, on average, are significantly higher in TNG100 than in MaNGA ($\approx50\,\kms$). We study the radial profiles of \emph{in-situ} and \emph{ex-situ} stars in TNG100 and discuss the extent to which each population contributes to the observed MaNGA profiles. Our analysis lends significant support to the idea that high-mass ($M_*\gtrsim10^{11}\,\msun$) ETGs in the present-day Universe are the result of a merger-driven evolution marked by major mergers that tend to homogenise the stellar populations of the progenitors in the merger remnant.
\end{abstract}




\section{Introduction}

In the standard cosmological framework, the formation and evolution of galaxies  is thought to be driven by mergers and the accretion of material from the intergalactic medium \citep[e.g.][]{Cimatti2019}.  In particular, for the assembly of massive early-type galaxies (ETGs), a \emph{two-phase formation scenario} has been proposed \citep[e.g.,][]{Naab2009ApJl,Oser2010ApJ,Hilz2013MNRAS}. In the first phase of this formation process ($z\gtrsim2$), ETGs are built from stars formed \emph{in situ}, i.e.\ within the same galaxy, while later, as a consequence of minor and major mergers, ETGs grow mainly by the accretion of stars formed \emph{ex situ}, i.e.\ in other galaxies.

A natural outcome of mergers experienced by ETGs is the evolution of scaling relations \citep[e.g.][]{Cimatti2019}, i.e.\ the observed empirical correlations between global galaxy properties, such as those relating luminosity (or mass) with stellar velocity dispersion \citep{FaberJackson1976ApJ}, size \citep{Kormendy1977ApJ}, or both (the so-called fundamental plane; \citealt{DjorgovskiDavis1987ApJ,Dressler1987ApJ}). The improvement of instrumentation technology together with increasing statistics in  recent surveys has enabled studies of those relations at different redshifts. Indeed, massive ETGs at high redshift are found to be compact, with an effective radius $R_\mathrm{e}$ smaller than that of galaxies of similar stellar mass in the present-day Universe \citep[e.g.,][]{Ferguson2004ApJ,vanderWel2014ApJ,Damjanov2019ApJ}.
Also the stellar mass--central velocity dispersion relation ($M_*{-}\sigma_\mathrm{e}$) evolves: on average, for a given stellar mass, the lower the redshift, the lower the velocity dispersion \citep[e.g.,][]{vandeSande2013ApJ,Belli2014ApJ,Belli2017ApJ,Tanaka2019ApJ,CSN2020MNRAS}.

Mergers and accretion not only affect global galaxy properties, but also the internal distributions of stellar properties. The spatial distributions of metallicity, chemical abundances, age and other properties of stellar populations in a galaxy enclose information on the evolutionary processes that have occurred across cosmic time. One possible way to investigate how progenitor stellar populations come together to form present day ETGs is to perform galaxy-scale high-resolution simulations, an approach recently adopted,  for instance, by \citet{Nipoti2020MNRAS} to study the 
effect of mergers on the internal distributions of the stellar initial mass function (IMF)
and the velocity dispersion in ETGs.
A second approach is the study of the evolution of stellar population properties using hydrodynamical cosmological simulations. For example, \citet[][]{Oser2012ApJ}, exploiting a set of fourty zoom-in hydrodynamical simulations of individual halos presented in \citet{Oser2010ApJ}, and \citet{RodriguezGomez2016MNRAS} using the Illustris \citep{Vogelsberger2014Nature,Vogelsberger2014MNRAS,Genel2014MNRAS,Sijacki2015MNRAS} simulations, studied the radial  distributions of in-situ and ex-situ stars in ETGs. With a similar approach, \citet{Barber2019MNRAS} studied IMF radial gradients in ETGs drawn from the Evolution and Assembly of GaLaxies and their Environments \citep[EAGLE;][]{Schaye2015MNRAS,Craine2015MNRAS,McAlpine2016A&C} cosmological simulations.


The presence of initial gradients in stellar metallicity are thoughts to be established during the first episodes of star formation \citep[e.g.,][]{Larson1974MNRAS,Thomas2005ApJ} with metallicity profiles that decrease towards the external regions of galaxies, but with fairly flat stellar age profiles.
As shown in \citet{Hirschmann2015MNRAS} using a set of ten high-resolution cosmological zoom simulations presented in \citet{Hirschmann2013MNRAS}, as well as in \citet{Cook2016ApJ} taking data from Illustris, the large number of mergers and interactions that occur in galaxies then tend to flatten metallicity profiles and almost flatten (or lead to slightly positive) age gradients both because of the mixing of stars with different metallicities and the accumulation of old stellar populations in the outer regions.
On the contrary, galaxies with few mergers may retain their original negative metallicity profiles with metal-poor regions dominating in the outskirts of galaxies \citep[e.g., ][]{Kobayashi2004MNRAS,Pipino2010MNRAS,TaylorKobayashi2017MNRAS}.
 
In the last few decades, integral field spectroscopy (IFS) has formed the basis of many surveys: SAURON \citep[Spectroscopic Areal Unit for Research on Optical Nebulae; ][]{Bacon2001MNRAS,deZeeuw2002MNRAS}, ATLAS$^\mathrm{3D}$ \citep{Cappellari2011MNRAS}, CALIFA \citep[Calar Alto Legacy Integral Field Array survey; ][]{Sanchez2016A&A},  SAMI \citep[Sydney-Australian-Astronomical-Observatory Multi-object Integral-Field Spectrograph][]{Croom2012MNRAS,Bryant2015MNRAS}, MASSIVE \citep{Ma2014ApJ} and MaNGA \citep[Mapping Nearby Galaxies at Apache Point Observatory; ][]{Bundy2015ApJ}.
These spatially-resolved surveys allow in depth studies of the properties of stellar populations in individual objects, therefore not limiting analyses only to the study of gradients, but revealing the 2D spatial distribution over the entire galaxy on the plane of the sky.
By analysing a set of ETGs with $\log(\mstar/\msun)>10.3$ in SAURON, \citet{Kuntschner2010MNRAS} found that stellar metallicity gradients become shallower with increasing stellar mass, while stellar age gradients are independent of stellar mass. \citet{Li2018MNRAS} using MaNGA galaxies with  $9<\log(\mstar/\msun)<12.3$ found metallicity gradients consistent with those of \citet{Kuntschner2010MNRAS}. Moreover,  \citet{Li2018MNRAS} found that stellar metallicity gradients show a strong dependence on stellar velocity dispersion: they peak (being most negative) at velocity dispersions of around $100\,\kms$. This radial dependence can be interpreted in terms of different evolutionary scenario for galaxies with different velocity dispersions. In particular, metallicity gradients tend to flatten at high velocity dispersions, perhaps indicating the rising role of mergers that redistribute stellar populations in these galaxies.

However, studies conducted so far that involve IFS surveys are also sometimes in disagreement. 
For example, \citet{Goddard2017bMNRAS} selected ETGs from MaNGA with $9<\log(\mstar/\msun)<11.5$. Although the galaxies were drawn from the same survey as used by \citet{Li2018MNRAS}, the authors derived metallicity profiles that become steeper towards higher masses. A similar result was found by \citet{Zheng2017MNRAS}, for ETGs in the MaNGA survey with $8.5<\log(\mstar/\msun)<11.5$.
In \citet[][]{Greene2015ApJ}, subsequently extended in \citet[][]{Greene2019ApJ} to larger radii, ETGs with $\log(\mstar/\msun)>11.6$ show shallow metallicity gradients and radius-independent age and $\alpha$-element abundances relative to iron, i.e. $\left[\alpha/\mathrm{Fe}\right]$.
By analysing a sample of 96 passive brightest cluster galaxies (BCGs) from the SAMI survey, \citet{Santucci2020ApJ} found negative metallicity gradients that tend to become shallower as the stellar mass increases, slightly positive age gradients and almost zero $\left[\alpha/\mathrm{Fe}\right]$ gradients, the latter tending to become slightly more negative with increasing mass. This study also revealed there to be no significant differences in the stellar profiles of the analysed properties between central and satellite galaxies, both at fixed stellar mass and as a function of halo mass, suggesting that the two galaxy populations follow a similar formation scenario, which appears to be independent of the environment.
Differences among these various studies (also when using the same galaxy survey) appear to result from a combination of different selection criteria adopted to identify ETGs, the stellar mass ranges considered, and the methods used to retrieve properties and their profiles. In addition, we have found that the radial range adopted to measure the gradients and whether the profiles are stacked in physical units or in units of $\reff$, can lead to some of these discrepancies.

Despite the relatively large number of IFS surveys, understanding whether a stellar population in a galaxy either formed in situ or was accreted from another progenitor is not a trivial task.
\citet{Oyarzun2019ApJ}, analysing more than 1000 ETGs with $10<\log(\mstar/\msun)<12$ from the MaNGA survey, studied the radial distributions of metallicity adopting three different stellar fitting codes, i.e. \textsc{FIREFLY} \citep{Wilkinson2017MNRAS,Comparat2017arXiv,Goddard2017MNRAS,Maraston2011MNRAS,Maraston2020MNRAS}, \textsc{Prospector} \citep{Leja2017ApJ,Johnson2019ascl} and \textsc{pPXF} \citep{CappellariEmsellen2004PASP,Cappellari2017MNRAS}.
As the mass increases, the flattening in the metallicity profiles was found to become more prominent at $R\gtrsim R_\mathrm{e}$. \citet{Oyarzun2019ApJ} interpreted this flattening using a toy model in which they assume that the low-mass tail of galaxies in their sample are representative of galaxies mainly constituted by stars formed in situ. For  high-mass galaxies, the inner part of the profiles ($R\lesssim R_\mathrm{e}$) is associated with an in-situ stellar population, while the external parts are considered to be dominated by stars accreted from other galaxies. Quantitatively, they infer the contribution of ex-situ stars within $R\approx2R_\mathrm{e}$ to be $\approx20\%$ of the total stellar mass in ETGs with $\log(\mstar/\msun)<10.5$, while in ETGs with $\log(\mstar/\msun)>11.5$ this fraction reaches $\approx80\%$ (consistent results are also presented in the observational works of \citealt{Edwards2020MNRAS} and \citealt{Davison2021MNRAS}). 

An alternative approach that allows to combine observations and simulations has been recently proposed by \citet{Nanni2022MNRAS}. In that paper, the authors built iMaNGA, a MaNGA-like galaxy sample considering both early- and late-type galaxies extracted from the cosmological simulation TNG50 \citep[][]{Nelson2019MNRAS,Pillepich2019MNRAS}.
Specifically, \citet{Nanni2022MNRAS}, collecting simulated galaxies from the snapshots between $z=0.01$ and $z=0.15$, so that covering the whole MaNGA redshift range, took into account all the instrumental effects and methods employed to acquire data for MaNGA sources. The specific use of TNG50 allowed the authors to take all the advantages of high-spatial-resolution data, and generate corresponding mock galaxy spectra. Along similar lines, \citet{Bottrell2022MNRAS}  presented RealSim-IFS5, a generalised tool for forward-modelling realistic synthetic IFS observations from hydrodynamical simulations. RealSim-IFS is able to reproduce cubes similar to those produced by the MaNGA survey Data Reduction Pipeline. Furthermore, extracting around 900 galaxies with $\log(\mstar/\msun)>10$ from TNG50, \citet{Bottrell2022MNRAS} applied RealSim-IFS to generate a synthetic MaNGA stellar kinematic survey.

In this work, we propose a physically-grounded model to provide an interpretative scenario for the radial distributions of stellar properties in observed ETGs in terms of in-situ and ex-situ stellar components. In particular, we compare the radial profiles of stellar mass surface density, metallicity, age and line-of-sight velocity dispersion of observed galaxies drawn from the data release 15 of the MaNGA survey with those of simulated ETGs extracted from the TNG100 simulation of the \textit{The Next Generation} Illustris project \citep[IllustrisTNG\footnote{Official website at \url{https://www.tng-project.org}.};][]{Springel2018MNRAS,Nelson2018MNRAS,Pillepich2018MNRAS,Naiman2018MNRAS,Marinacci2018MNRAS}. Simulated galaxies are broken down into in-situ and ex-situ stellar populations using the methods presented in \citet{RodriguezGomez2015MNRAS} which serves as a prediction for the gradients of these two populations in massive ETGs. Indeed, the main scope of this work is to suggest a possible evolutionary scenario of the underlying hierarchical stellar mass assembly history of present-day ETGs. With this goal in mind, we focus the analysis on the study of radial distributions of the above-mentioned stellar properties for both MaNGA and IllustrisTNG galaxies, considering quantities \emph{at face value}, i.e. as those directly derived from the pipelines and stellar fitting codes for MaNGA, and those from the TNG100 simulations.

This paper is organised as follows. In \autoref{sec:manga_tng_samples} we describe the galaxy samples and the criteria adopted to select ETGs. The method used to compute radial profiles from simulations is described in \autoref{sec:methods_manga_tng}. Our results are presented in \autoref{sec:manga_tng_results}.
In \autoref{sec:discussion} we discuss the implications of the analysis and compare with previous studied, while
\hyperref[sec:conclusion_manga_tng]{section~\ref*{sec:conclusion_manga_tng}} presents our conclusions.

Throughout this paper, we assume a $\Lambda$CDM cosmological framework with cosmological parameters derived from \citet{Planck2016AAP}, i.e.  $\Omega_{\Lambda,0}=0.6911$, $\Omega_{m,0}=0.3089$, $\Omega_{b,0}=0.0486$, and $H_0=67.74\,\kms\mathrm{Mpc}^{-1}$. 

\section{Observed and simulated galaxy samples}
\label{sec:manga_tng_samples}

In this section, we describe the selection criteria and physical properties of the observed (MaNGA) and simulated (TNG100) samples and the methods adopted to compare the two samples.

\subsection{The MaNGA survey}
\label{ssec:manga_survey}

The MaNGA survey \citep{Bundy2015ApJ,Yan2016AJ}, one of the three components of the fourth generation of SDSS \citep{York2000AJ,Gunn2006AJ,Blanton2017AJ} mapped with the 2.5 m telescope Apache Point Observatory $\approx10000$ galaxies with $\log(\mstar/\msun)>9$ in the redshift range $0.01\lesssim z\lesssim0.15$, providing spatially-resolved spectra for each source.
The galaxy sample is taken from an extended version of the original NASA-Sloan Atlas \citep[NSA {\ttfamily v1\_0\_1}\footnote{Available at \url{https://www.sdss.org/dr15/manga/manga-target-selection/nsa/}.};][]{Blanton2011AJ} catalogue. By exploiting the IFS technique \citep{Smee2013AJ,Drory2015AJ,Law2015AJ}, galaxies in MaNGA are observed with a set of 17 hexagonal bundles, each composed of fibers with a diameter that varies from $12^\dprime$ (with 19 fibers) to $32^\dprime$ (with 127 fibers). Each fiber has a diameter of $2^\dprime$.
MaNGA achieves a uniform radial coverage of galaxies to $1.5\,R_\mathrm{e}$ and $2.5\,R_\mathrm{e}$, for $\approx 2/3$ (\textit{Primary Sample}) and $\approx 1/3$ (\textit{Secondary Sample}) of the final sample. The observations provide a wavelength coverage in the range $3600{-}10300\,\angstrom$, with a spectral resolution of $R\sim1400$ at $\lambda\sim4000\,\angstrom$ and $R\sim2600$ at $\lambda\sim9000\, \angstrom$ \citep[see][]{Smee2013AJ}.

The MaNGA observations used in this work were previously reduced by the Data Reduction Pipeline \citep[DRP;][]{Law2016AJ,Yan2016bAJ}. 
Both the de-projected distances and stellar kinematic maps are computed using the Data Analysis Pipeline \citep[DAP;][]{Westfall2019AJ} for MaNGA.
The MaNGA galaxies forming our observed sample are taken from the SDSS data release 15 \citep[DR15, hereafter simply MaNGA;][]{Aguado2019ApJS} which corresponds to the first 4675 observed MaNGA galaxies\footnote{Available at \url{https://www.sdss.org/dr15/manga/manga-data/}.}. 

To study the behaviour of radial profiles of observed ETGs we use measurements of stellar mass surface density, metallicity, and age derived from two full spectral fitting codes: \textsc{FIREFLY}\footnote{Available at \url{https://github.com/FireflySpectra/firefly_release}.} \citep{Wilkinson2017MNRAS,Comparat2017arXiv,Goddard2017MNRAS,Maraston2011MNRAS,Maraston2020MNRAS} and \textsc{Prospector}\footnote{Available at \url{https://github.com/bd-j/Prospector}.} \citep{Leja2017ApJ,Johnson2019ascl}. The use of two different methods will help to quantify the presence of systematic biases caused by different assumptions, priors and fitting methods \citep{Conroy2013ARA&A}. In addition, we take into account estimates of line-of-sight stellar velocity dispersion obtained by using the \textsc{pPXF} code\footnote{Available at \url{http://www-astro.physics.ox.ac.uk/~mxc/software/}} \citep{CappellariEmsellen2004PASP,Cappellari2017MNRAS}. In the following we briefly summarise the settings adopted for the three stellar population fitting codes.
\begin{itemize}
\item \textsc{FIREFLY} (Fitting IteRativEly For Likelihood analYsis) is a $\chi^2$-minimisation fitting code for deriving the stellar population properties. This code aims at disentangling stars and dust, subtracting the low-order continuum shape before fitting spectra. A set of simple stellar populations (SSPs) with a variety of age and metallicity are considered iteratively, in order to minimise the $\chi^2$ fitting procedure, allowing \textsc{FIREFLY} to fit non-parametric star formation histories (SFHs). We adopt the stellar population models of \citet{Maraston2011MNRAS}, the MILES stellar library \citep{SanchezBlazquez2006MNRAS,Vazdekis2010MNRAS}, and a \citet{Chabrier2003PASP} IMF.
The set of SSPs used are spread over the range 6.5 Myr${-}$15 Gyr in age, while metallicity can assume values in the range $-2.3\leq\log(Z_*/Z_\odot)\leq0.3$. 
The wavelength range covered by the library is $4000{-}7400\,\angstrom$. We include only spectra with $S/N>10$ \citep[see][]{Wilkinson2017MNRAS,Goddard2017bMNRAS}, and we mask emission lines.

\item \textsc{Prospector} is a code able to infer stellar population properties from photometric and/or spectroscopic data with flexible models. It is based on the original stellar population synthesis code \textsc{FSPS}\footnote{Available at \url{https://github.com/cconroy20/fsps}.} \citep{Conroy2009ApJ,Conroy2010ApJ}. \textsc{Prospector} provides the posterior distribution of a stellar population parameter space (externally defined by users), uncertainties, and degeneracies. We adopt the MILES stellar population library, the MIST isochrones  \citep{Dotter2016ApJS,Choi2016ApJ} and a \citet{Kroupa2001MNRAS} IMF\footnote{For our purpose, the assumption of a Kroupa IMF or a Chabrier IMF to retrieve stellar population properties is almost indistinguishable.}. The fitting procedure explores a ten-dimensional parameter space. In this fit, the dust optical depth in the $V$-band, stellar mass, stellar velocity dispersion, and mass-weighted metallicities are taken into account. Moreover, non-parametric SFHs with a continuity prior are considered. Following the same approach described in \citet{Leja2019ApJ}, our parameter space considers the star formation rate (SFR) spanning the following time intervals:
      $0\!<\! t\! <\!  30\,\mathrm{Myr},\,
      30\,\mathrm{Myr}\!<\! t\! <\! 100\,\mathrm{Myr},\,
     100\,\mathrm{Myr}\!<\! t\! <\! 330\,\mathrm{Myr},\,
     330\,\mathrm{Myr}\!<\! t\! <\! 1.1\,\mathrm{Gyr},\,
     1.1\,\mathrm{Gyr}\!<\! t\! <\! 3.6\,\mathrm{Gyr},\,
     3.6\,\mathrm{Gyr}\!<\! t\! <\!11.7\,\mathrm{Gyr}$ and 
   $11.7\,\mathrm{Gyr}\!<\! t\! <\!13.7\,\mathrm{Gyr}$.
The priors used for our \textsc{Prospector} runs are listed in \autoref{tab:prospector}. Finally, the posterior distributions are obtained exploiting the Dynamic Nested Sampling package {\ttfamily dynesty} \citep{Speagle2020MNRAS}.

\begin{table}
\centering
\caption{List of priors used for our \textsc{Prospector} runs. Column 1: parameter. Column 2: prior.}
\begin{tabular}{lc}
\toprule
\toprule
\addlinespace
Parameter & Prior \\
\addlinespace
\midrule
Star formation history & Continuity\\
\addlinespace
dust2 & TopHat $(0,1)$ \\
\addlinespace
Stellar metallicity $\log Z_*\,[\mathrm{Z}_\odot]$ &  TopHat $(-2,\,0.3)$\\
\addlinespace
Formed stellar mass $M_*/\msun$ & LogUniform($10^7,\,10^{12}$) \\
\addlinespace
Velocity dispersion $\sigma_*$ [$\kms$]   & TopHat $(0.1,\,400)$ \\
\addlinespace
\bottomrule
\bottomrule
\end{tabular}
\label{tab:prospector}
\end{table}

\item The Penalized Pixel-Fitting method (\textsc{pPXF}) code derives the stellar or gas kinematics and stellar population from absorption-line spectra of galaxies, using a maximum penalized likelihood method. The original approach was presented in \citet{CappellariEmsellen2004PASP} and then improved in \citet{Cappellari2017MNRAS}. We used \textsc{pPXF} to estimate line-of-sight velocity dispersions for our observed ETGs. 
The penalisation of pixels that are not well fit minimises the mismatch with the templates employed. We ran \textsc{pPXF} with the MILES library.
\end{itemize}

\subsection{TNG100 simulation}
\label{ssec:IllustrisTNG}
In this work, we extract simulated ETGs from IllustrisTNG\footnote{Official website at \url{https://www.tng-project.org}.} \citep{Springel2018MNRAS,Nelson2018MNRAS,Pillepich2018MNRAS,Naiman2018MNRAS,Marinacci2018MNRAS}, the successor to the original Illustris\footnote{Official website at \url{https://www.illustris-project.org}.} simulation suite \citep{Vogelsberger2014MNRAS,Vogelsberger2014Nature,Genel2014MNRAS,Sijacki2015MNRAS}. 
The data are publicly available\footnote{https: //www.illustris-project.org/data/} and presented in \citet{NelsonMNRAS2021}.
IllustrisTNG is a state-of-the art magneto-hydrodynamic cosmological simulation that models the formation and evolution of galaxies within the $\Lambda$CDM framework. As its predecessor, IllustrisTNG exploits all the advantages of the \textit{unstructured moving-mesh} hydrodynamic method \textsc{arepo} \citep{Springel2010MNRAS}, but improves the numerical methods, the subgrid physical model, and the recipe for galaxy feedback both from stars and AGN. In particular, IllustrisTNG is equipped with a novel dual mode (thermal and kinetic) AGN feedback that shapes and regulates the stellar component within massive systems, maintaining a realistic gas fraction \citep{Weinberger2017MNRAS}. Also the feedback model from galactic winds has been improved to have better representation of low- and intermediate-mass galaxies \citep{Pillepich2018aMNRAS}.

The IllustrisTNG model was calibrated to significantly reduce tensions between the original Illustris suite and observations.
As shown in Figure 4 of \citet{Pillepich2018aMNRAS}, some of these relevant improvements include the star formation rate density as a function of time, the stellar-to-halo mass relation in the present-day Universe, the stellar mass function, the black hole mass${-}$stellar mass relation, the black hole mass${-}$halo mass relation, the gas content within virial radii, and galaxy sizes.

The IllustrisTNG simulation suite consists of three simulation volumes: TNG50 \citep{Nelson2019MNRAS,Pillepich2019MNRAS}, TNG100 and TNG300, corresponding to three different box sizes with sides of about $50\,\mpc$, $100\,\,\mpc$ and $300\,\mpc$, respectively.
The project assumes a $\Lambda$CDM cosmology with cosmological parameters taken from \citet{Planck2016AAP}. Each run starts at $z=127$ using the Zeldovich approximation and evolves down to $z=0$.

We use the highest-resolution version of the medium volume size TNG100, i.e. TNG100-1 (hereafter, simply TNG100). This run includes approximately $2\times1820^3$ resolution elements. The dark matter (DM) and baryonic mass resolutions are $m_\mathrm{DM}=7.5\times10^6\,\msun$ and $m_\mathrm{b}=1.4\times10^6\,\msun$. The softening length employed for this version for both the DM and stellar components is $\epsilon=0.74\,\kpc$, while an adaptive gas gravitational softening is used, with a minimum $\epsilon_\mathrm{gas,min}=0.185\,\mathrm{kpc}$. In particular, we take into account the properties of subhalos from snapshot $\#91$, corresponding to $z=0.1$, close to the mean redshift of galaxies in the MaNGA survey.

\subsubsection{In-situ \& ex-situ stars in IllustrisTNG galaxies}

In the last decade, cosmological simulations have suggested that accretion contributes to the mass and size growth of massive galaxies. The fraction of accreted stars depends on the stellar and DM masses \citep[e.g.,][]{Oser2010ApJ,Lackner2012MNRAS,Pillepich2014MNRAS}.
From the original Illustris simulation suite, \citet{RodriguezGomez2016MNRAS} derived the ex-situ fraction for galaxies with stellar masses between $10^9\,\msun$ and $10^{12}\,\msun$. They found that this fraction increases from $\lesssim10\%$ in the least massive galaxies to above $80\%$ in the most massive systems. A similar analysis has been conducted on the IllustrisTNG runs: \citet{Pillepich2018MNRAS} analysed stellar masses within different apertures and found that, at $z=0$, the low-mass tail of galaxies are mainly formed by in-situ stellar particles, while central galaxies living in the most massive halos, i.e. $\log(M_{200\mathrm{c}}/\msun)>14$, accreted more than $80\%$ of their total stellar mass via mergers. Moreover, by considering stellar masses within an aperture larger than $100\,\kpc$, the ex-situ fraction is found to be dramatically dominant at these distances, exceeding sometimes $90\%$ of the total mass. The relative contribution in massive systems of the ex-situ component reaches around $60\%$ in the innermost regions ($<10\,\kpc$).

In this paper, we adopt the same definition of in-situ and ex-situ stars as used in  \citet{RodriguezGomez2016MNRAS}, \citet{Pillepich2018MNRAS}, and \citet{Tacchella2019MNRAS}, exploiting the method for reconstructing the baryonic merger trees of \citet{RodriguezGomez2015MNRAS}:
\begin{itemize}
\item \emph{in-situ stars} are those stellar particles that formed in a galaxy belonging to the main progenitor branch of the merger tree.
\item \emph{ex-situ stars} are those stellar particles that, at the time of their formation, were bound to a galaxy outside the main progenitor branch of the descendant galaxy.
\end{itemize}




\subsection{ETG selection}
\label{ssec:ETG_selection}

We aim to perform a comparison that is as consistent as possible between TNG100 and MaNGA. The first question at hand is how to select ETGs.  We opt for a  simple selection based on $(g-r)$ rest-frame colours, identifying hereafter ETGs as \textit{Red Galaxies}, that is those galaxies with $(g-r)>0.6$, a value that marks the transition between the blue cloud and the red sequence of galaxies. 
\citet{Nelson2018MNRAS} consider three models for assigning $ugriz$ magnitudes to simulated galaxies, comparing them with those of SDSS galaxies. The authors discuss in detail the results obtained using colours derived through the so-called \textit{resolved dust model} (Model C) which accounts for the presence of dust, following the distribution of neutral gas in galaxies, and adding also the attenuation caused by the presence of metals. The data from Model C are available in the supplementary catalogue \textsc{SDSS Photometry, colours, and Mock Fiber Spectra}\footnote{Available at \url{https://www.tng-project.org/data/downloads/TNG100-1/}.}. In \citet{Nelson2018MNRAS}, the IllustrisTNG colours were compared with the observed colours of SDSS DR12 \citep{Alam2015ApJS} galaxies in the present-day Universe ($z<0.1$). The distributions of $(g-r)$ colours recovers the colour bimodality of SDSS galaxies. In SDSS, the blue and red simulated galaxy populations show two characteristic peaks at $(g-r)\approx0.4$ and $\approx0.8$, respectively. Moreover, above $M_*\simeq10^{10.5}\,\msun$ the colour bimodality tends to disappear and red galaxies dominate.
For this paper, to select ETGs in our observed MaNGA sample, we retrieved the $ugriz$ Petrosian magnitudes from the NSA catalogue.
To sum up, in the following, we will present the results of our analysis for MaNGA and TNG100 ETGs selected as such considering only objects with $(g-r)>0.6$.

\subsection{Stellar mass estimates}
\label{ssec:Stellar_mass_estimates}

Another question to consider when comparing observations and simulations is the consistency of the stellar mass measurements. Indeed, differences in the measurements of stellar masses can be caused by several factors, such as the fitting method used to derive luminosities and colours, as well as the stellar population synthesis models and libraries assumed.  \citet{Sonnenfeld2019A&A} discuss the differences in deriving luminosity of galaxies in massive ETGs observed with the Hyper-Supreme Cam \citep[HSC;][]{Miy++18} Subaru Strategic Program \citep[][DR1]{Aihara2018PASJ}, assuming either a simple Sérsic fit or a Sérsic+Exponential fit. The difference between the two methods can cause a variation of around $0.1\,\mathrm{dex}$ on the measurements of luminosity for the same object.
Moreover, a different assumption of IMF can imply a global shift of stellar masses and the potential presence of IMF radial variations can introduce biases. The radius used to estimate stellar mass, or the quality of the observational data, can also be an important factor \citep[e.g.,][]{Huang2018, Ardila2021MNRAS}.

In this work we assume the stellar mass estimates from the \textsc{UPenn\_PhotDec\_MsSTAR}\footnote{Available at \url{http://alan-meert-website-aws.s3-website-us-east-1.amazonaws.com/fit_catalog/download/index.html}.} catalogue of \citet{Meert2015MNRAS} for MaNGA ETGs.
In particular, these stellar masses, obtained by multiplying the stellar mass-to-light ratios ($M_*/L$) from \citet{Mendel2014ApJS} by the luminosities from the PyMorph SerExp (Sérsic+Exponential) photometry, assume $M_*/L$ fitting models that account for the effects of dust extinction (Table 3 of \citealt{Mendel2014ApJS}). 

For TNG100 galaxies we consider the $2\mathrm{D}$ projected stellar mass defined as the sum of all bound stellar particles within a projected radius $R=2R_\mathrm{hm}$,  where $R_\mathrm{hm}$ is the radius of a circle containing half of all stellar particles bound to each subhalo.
Hereafter, we will refer to \emph{2hmr mass} as the projected mass within a radius of $R=2R_\mathrm{hm}$.

\begin{figure*}
    \centering
    \includegraphics[width=2\columnwidth]{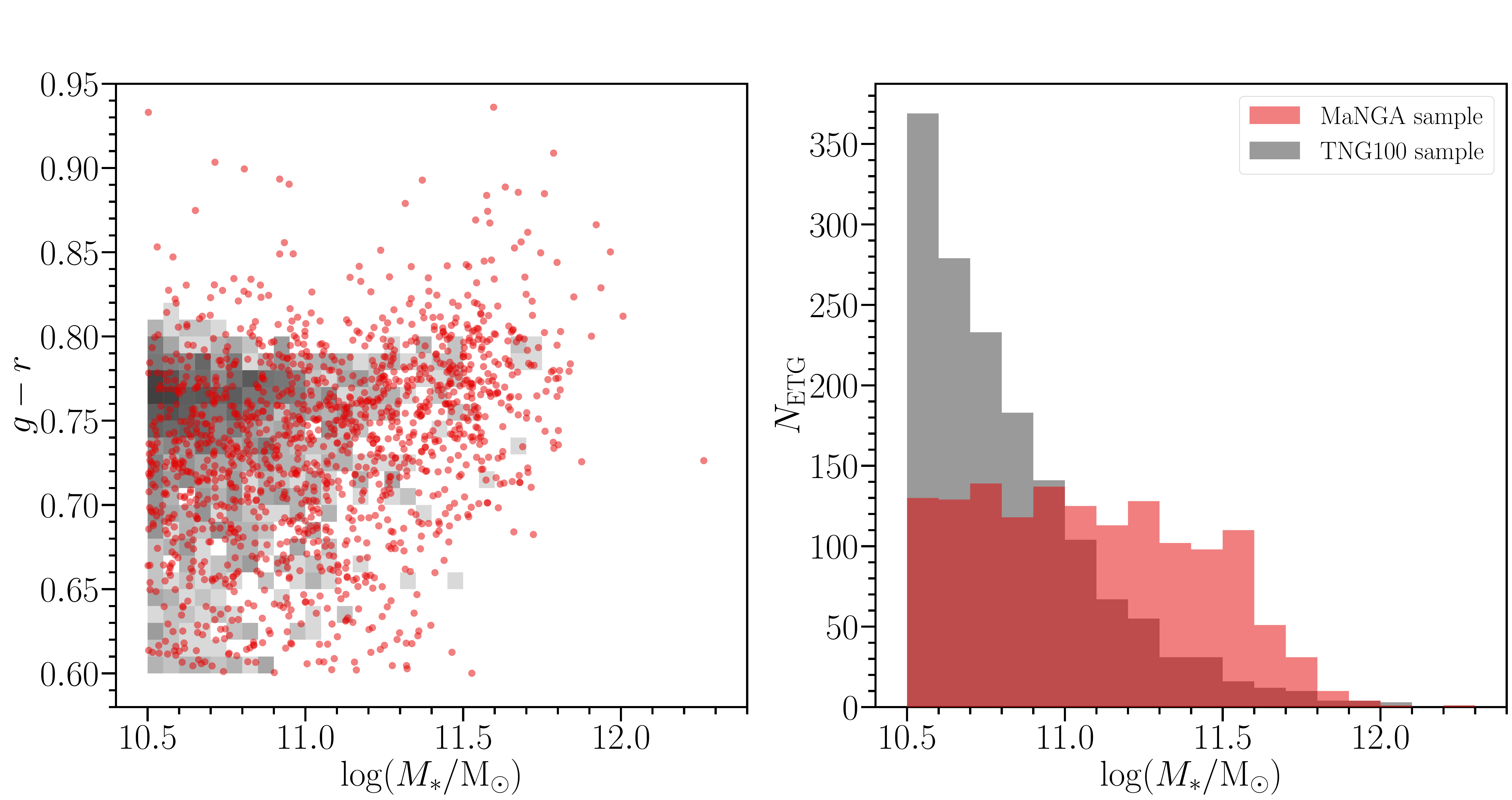}
    \caption{Left: MaNGA (red dots) and TNG100 (2D grey histogram) ETGs. Right panel: 1D histograms of the mass distribution for the MaNGA (red histogram) and the TNG100 (grey histogram) estimates. 
    The histogram of MaNGA ETGs stellar masses is flat due to the MaNGA selection function \citep[see][]{Wake2017AJ}.}
    \label{fig:manga_illustristng_samples}
\end{figure*}

\autoref{tab:manga_illustristng_samples} summarises the properties of the final TNG100 and MaNGA \emph{Red Galaxy samples}, i.e. objects with $\log(M_*/\msun)\geq10.5$ and $(g-r)>0.6$. \autoref{fig:manga_illustristng_samples} shows the colour--mass diagrams and the stellar mass distributions of both samples.
As clearly visible,
the distribution of the stellar masses for MaNGA galaxies appears almost flat. The reason for such distribution is due to the original MaNGA sample design: as argued in \citet{Wake2017AJ}, the MaNGA sample was built in such a way that the most massive galaxies are located at higher redshifts, but, at the same time, both the Primary and Secondary samples are selected to have flat stellar mass distributions. We will account for this effect later in the paper.

\begin{table*}
\centering
\caption{Summary table of the MaNGA and TNG100 samples. Column 1: sample. Column 2: number of ETGs. Column 3: stellar mass range. Column 4: mean stellar mass. Column 5: median stellar mass. Stellar masses are in units of $\msun$.}
\begin{tabular}{lcccc}
\toprule
\toprule
\addlinespace
Sample & $N_\mathrm{ETG}$ &
($\log M_{*,\mathrm{min}}$;
 $\log M_{*,\mathrm{max}}$) &
$ \log M_\mathrm{*,mean}$ & $\log M_\mathrm{*,median}$ \\
\addlinespace
\midrule
\addlinespace
MaNGA & 1427 & (10.50; 12.26) & 11.07 & 11.04 \\
\addlinespace
TNG100 & 1543 & (10.50; 12.27) & 10.83 & 10.76 \\
\addlinespace
\bottomrule
\bottomrule
\end{tabular}
\label{tab:manga_illustristng_samples}
\end{table*}

\section{Radial profiles of stellar properties}
\label{sec:methods_manga_tng}

In \autoref{ssec:radial_profiles_manga} and \autoref{ssec:radial_profiles_tng} we describe the method used to compute the radial profiles from the observed and simulated 2D stellar galaxy images.
In \autoref{ssec:mass_bins} and
\autoref{ssec:median_profiles} we describe how we compute median profiles in different stellar mass bins.

\subsection{Radial profiles for MaNGA ETGs}
\label{ssec:radial_profiles_manga}

To obtain stellar properties at different galactocentric distances for each MaNGA galaxy we adopt the approach described in \citet{Oyarzun2019ApJ}, with the difference that we consider radial binning in physical units instead of units of effective radii. 
Specifically, by considering the axis ratio of each source obtained from $r$-band photometric images, elliptical polar radii are associated to spaxels. We then bin in five concentric elliptical annuli each galaxy map, assuming the following radii as the edges of each bin: $R/\mathrm{kpc}=\{0;\,2;\,4;\,10;\,20;\,100\}$.
The choice of using a radial binning in physical units is justified by the fact that effective radius measurements may be affected by the depth of the survey. For example, HSC measures different effective radii than SDSS \citep[][]{Huang2018}.

The next step after radial binning consists in shifting spectra to the rest-frame by taking the stellar systemic velocity from DAP as a reference. A Voronoi binning is then applied to the maps, considering a minimum $S/N=10$ in each bin. Spectra belonging to the same annulus are co-added and, after running \textsc{pPXF} with the MILES library, they are stacked to estimate the line-of-sight stellar mean velocity and velocity dispersion.

\subsection{Radial profiles for TNG100 ETGs}
\label{ssec:radial_profiles_tng}

To obtain the radial profiles of stellar properties for TNG100 galaxies, we apply the same method presented in \citet{Ardila2021MNRAS}.
We firstly project the 3D particle distributions of simulated galaxies on a 2D $X{-}Y$ plane using the {\ttfamily hydrotools} package \citep{Diemer2018,Diemer2019MNRAS}. For each subhalo, the 2D map consists of 300 pixels per side, with a resolution of $1\,\kpc$ per pixel (for a total physical side length of the map of $300\,\kpc$). 
To extract the 1D stellar profiles we then use the method presented in \citet{Huang2018} and also used by \citet{Ardila2021MNRAS}, which we now briefly describe. We extract 1D stellar mass surface density profiles using the galaxy surface brightness profile function included in the {\ttfamily kungpao} package\footnote{The {\ttfamily kungpao} library is available at \url{https://github.com/dr-guangtou/kungpao/}}.
Galaxy centroids are identified by means of {\ttfamily extract}, a function included in the {\ttfamily sep} library, and the {\ttfamily ellipse} algorithm is used to fit concentric elliptical isophotes. The position angle and ellipticity of these isophotes are the mean values from the 2D fitting procedure of the galaxy maps.
The isophotes are spread over the range $1{-}150\,\kpc$, in 20 concentric elliptical annuli of constant width in logarithmic space.
For stellar metallicity, age, and line-of-sight velocity dispersion profiles, we use the same centre and ellipticity of the isophotes computed on the stellar mass surface density mass maps.
To derive mass-weighted stellar metallicity, age, and velocity dispersion each pixel is weighted by the corresponding value of stellar mass in that pixel. 
The entire procedure is applied to both the in-situ and ex-situ stellar populations, starting from their 2D stellar property maps.

Another effect we accounted for is about the differences in spatial resolutions. 
For this, we associate each simulated galaxy an angular diameter distance, assuming a redshift drawn from the  $z{-}\mstar$ distribution of the MaNGA sample (see \autoref{fig:z_mass_planes} in  \hyperref[app:angular_distances]{Appendix A}). 
For each TNG100 ETG, we smooth the galaxy map with a 2D Gaussian filter kernel:
\begin{equation}\label{eq:kernel}
\sigma_{\mathrm{kernel},i}=\sqrt{\mathcal{R}_{\mathrm{MaNGA},i}^2-\mathcal{R}_\mathrm{TNG}^2},
\end{equation}
where $\mathcal{R}_\mathrm{TNG}=1\,\kpc$ is the resolution of the TNG100 sample, $\mathcal{R}_{\mathrm{MaNGA},i}=\sin(\mathrm{PSF}_\mathrm{MaNGA})d_{A,i}$, with $\mathrm{PSF}_\mathrm{MaNGA}=2.5^\dprime$ ($\simeq1.21\times10^{-5}$ in radians), and $d_{A,i}$ is angular diameter distance (in $\mathrm{kpc}$) for the $i$-th galaxy in the TNG100 sample determined as described above. For example, at $z=0.05$, $\mathcal{R}_{\mathrm{MaNGA}}\simeq2.52\,\mathrm{kpc}$.  For a given simulated ETG we compute two types of radial profiles for each stellar physical property (for both the in-situ and ex-situ stellar populations): the \emph{uncovolved profile} and the \emph{convolved profile}, the latter obtained by smoothing the projected maps with the $2\mathrm{D}$ Gaussian filter kernel $\sigma_\mathrm{kernel}$.

\subsection{Building stellar mass bins}
\label{ssec:mass_bins}

To compare the profiles of stellar properties between MaNGA and TNG100, we divide galaxies into bins of stellar mass and compute the median profile in each stellar mass bin along with the associated uncertainties.
As a fiducial choice, we compare galaxies at \emph{fixed stellar mass}. The three stellar mass bins used are $10.5\leq\log(\mstar/\msun)<11$, $11\leq\log(\mstar/\msun)<11.5$, and $\log(\mstar/\msun)\geq11.5$.
In \hyperref[app:number_density_cases]{Appendix B}, we present the same analysis using number-density-based bins, highlighting the differences with respect to the use of bins at fixed stellar mass.

\subsection{Building median radial profiles with errors}
\label{ssec:median_profiles}

We use a Bayesian hierarchical approach \citep[e.g.,][]{CSN2020MNRAS} to estimate the median values and the associated $1\sigma$ uncertainties on the observed and simulated radial profiles.  We assume that each stellar property $X$ in any radial bin has a Gaussian distribution, so that its likelihood can be written as
\begin{equation}
\mathrm{P}(X|X^\mathrm{data},\sigma_X^\mathrm{data},\mu,\sigma)=\frac{1}{\sqrt{2\pi\sigma_X^2}}\exp\left\{-\frac{(X^\mathrm{data}-\mu)^2}{2\sigma_X^2}\right\},
\label{eq:likelihood_manga}
\end{equation}
where $X$ is the quantity that we want to infer in each radial bin (e.g., the logarithm of the stellar mass surface density), while $X^\mathrm{data}$ and $\sigma_X^\mathrm{data}$ are the data values and their related uncertainties, respectively. The variance in \autoref{eq:likelihood_manga} has the form
\begin{equation}
\sigma_X^2={\sigma_X^\mathrm{data}}^2+\sigma^2. 
\label{eq:variance_likelihood_manga}
\end{equation}
In \autoref{eq:likelihood_manga} and in \autoref{eq:variance_likelihood_manga}, $\mu$ and $\sigma$ are the two \emph{hyper-parameters} of our Bayesian hierarchical approach and represent the mean value and the intrinsic scatter of the distribution of the quantity $X$, respectively.
We underline that, in the case of simulated ETGs, \autoref{eq:variance_likelihood_manga} reduces to $\sigma_X^2=\sigma^2$, since no uncertainties are associated to simulated properties.
These parameters are estimated independently in each bin.
In \autoref{tab:hyper_manga}, we list the priors adopted for each property.
\begin{table}
\centering
\caption{Hyper-parameters used to compute MaNGA and TNG100 profiles of stellar properties. Column 1: stellar property. Column 2: uniform prior on the mean (lower bound; upper bound). Column 3: uniform prior on the intrinsic scatter (lower bound; upper bound).}
\begin{tabular}{lcc}
\toprule
\toprule
\addlinespace
Stellar property &
$(\mu_\mathrm{min}; \mu_\mathrm{max})$ &
$(\sigma_\mathrm{min}; \sigma_\mathrm{max})$ \\
\addlinespace
\midrule
\addlinespace
Surface density $\log \Sigma_* \,[\msun\,\kpc^{-2}]$ & $(0;11)$ & $(0;2)$\\
\addlinespace
Metallicity $\log Z_*\,[\mathrm{Z}_\odot]$& $(-1;1)$ & $(0;1)$\\
\addlinespace
Age $\left[\mathrm{Gyr}\right]$& $(0;13)$ & $(0;5)$\\
\addlinespace
Velocity dispersion $\sigma_*\,[\kms]$& $(0;350)$ & $(0;100)$\\
\addlinespace
\bottomrule
\bottomrule
\end{tabular}
\label{tab:hyper_manga}
\end{table}
The stellar properties of MaNGA and TNG100 galaxies are sampled adopting a Markov chain Monte Carlo (MCMC) approach using  10 random walkers and 300 steps (removing the first 200 steps) for each run to reach the convergence of the hyper-parameter distributions. We use the Python adaptation of the affine-invariant ensemble sampler of \citet{GW2010}, \textsc{emcee} by \citet{Foreman-Mackey2013PASP}.

\subsection{MaNGA and TNG100 stellar properties}
In our comparison between MaNGA and TNG100 ETGs, we consider the circularised radial distributions of the stellar mass surface density $\Sigma_*$, stellar metallicity $Z_*$, stellar age, and line-of-sight stellar velocity dispersion $\sigma_*$.

\begin{figure*}
    \centering
    \includegraphics[width=.95\textwidth]{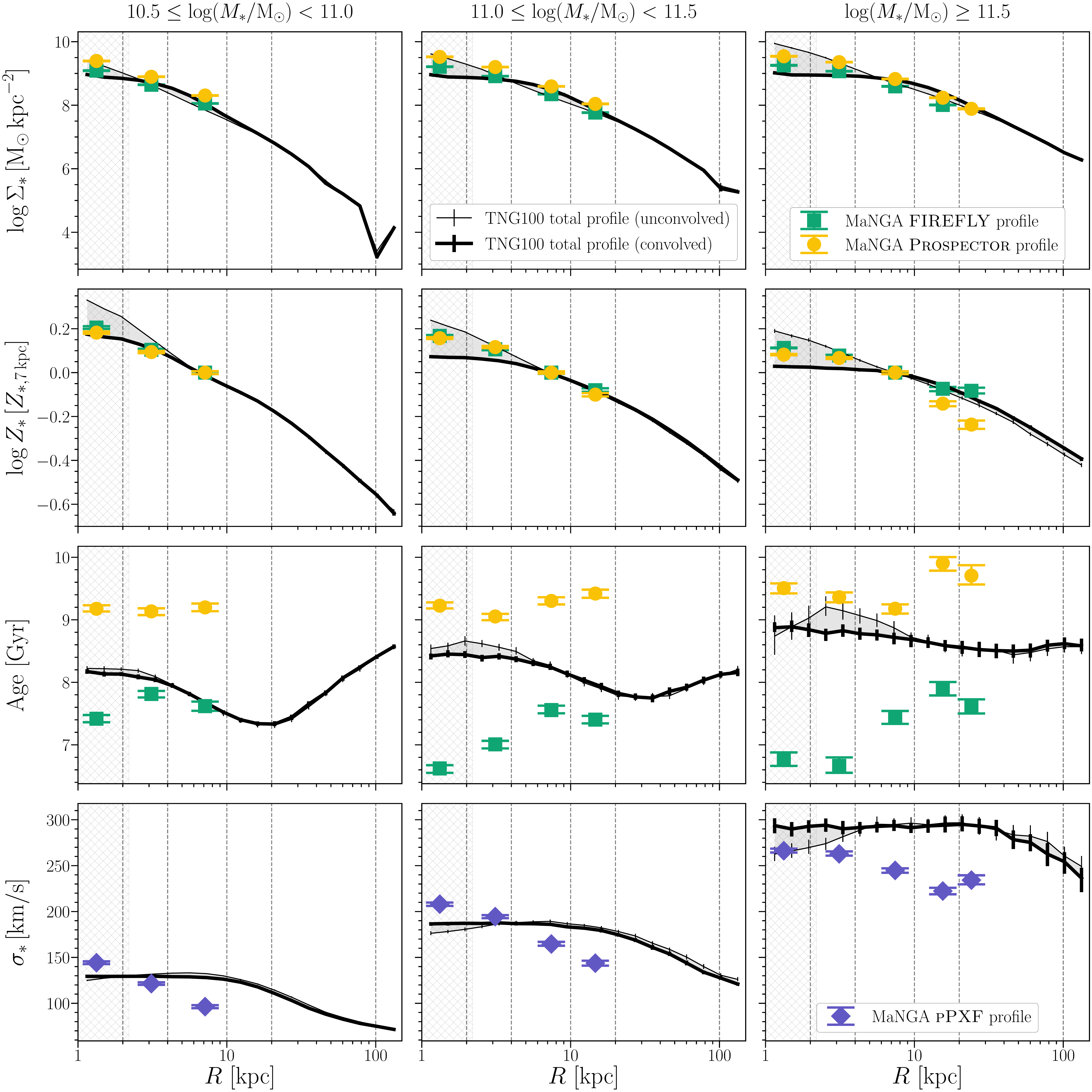}
        \caption{Radial profiles of  stellar mass surface density, metallicity (normalised by the corresponding values of metallicity at $\approx7\,\kpc$), age, and line-of-sight velocity dispersion (from top to the bottom) in three bins of stellar mass for MaNGA and TNG100 ETGs. Green, yellow and violet dots represent the median estimates respectively for MaNGA galaxies from \textsc{FIREFLY}, \textsc{Prospector} and \textsc{pPXF}. Both the stellar metallicity and age measurements in MaNGA and TNG100 ETGs are mass weighted. Velocity dispersions are luminosity weighted for MaNGA and mass weighted for TNG100 sources. The vertical grey dashed lines indicate the 5 radial bins for MaNGA.
        The two black curves represent the median values of each stellar property for the total stellar population in TNG100. The intrinsic profiles are shown with the thin curve whereas the thick curves indicate the results when convolved with the MaNGA PSF. The light grey hatched area ($R\lesssim2.1$ kpc) shows three times the gravitational softening length of the stellar particles in TNG100. The grey shaded area is the region that lies in between the profiles obtained from the original and the convolved TNG100 maps. Because we account for the MaNGA PSF but not for the effects of the resolution of the simulation, the grey shaded area gives a sense of the uncertainty in the comparison in the inner regions. The errorbars represent the $1\sigma$ uncertainties on the median for MaNGA and TNG100 estimates. }
    \label{fig:profiles_mass_dusty}
\end{figure*}

As already described in \autoref{ssec:radial_profiles_manga} and in \autoref{ssec:radial_profiles_tng}, in order to construct the radial profiles of the aforementioned properties, we build 5 concentric radial bins for MaNGA ETGs from the innermost regions out to $100\,\kpc$ and 20 log-spaced bins out to $150\,\kpc$ for TNG100 objects, for which we are able to split the relative contribution of in-situ and ex-situ stellar populations.
We make use of mass-weighted stellar metallicities and ages for both MaNGA and TNG100 galaxies. For these two properties, we also tested the luminosity-weighted measurements for MaNGA galaxies finding no significant difference between the two choices.
Velocity dispersions, instead, are mass weighted for simulated galaxies and luminosity-weighted for observed sources
\footnote{For TNG100 galaxies, the maps of stellar metallicity, age, and velocity dispersion are all weighted by stellar masses for consistency within the simulated sample. We note that, for the velocity dispersion, the comparison between simulated and observed galaxies is not fully self-consistent, because the velocity dispersion of the MaNGA galaxies is weighted by luminosity. However, as shown in \autoref{fig:profiles_mass_dusty}, the simulated galaxies have old stellar populations ($\gtrsim7\,\mathrm{Gyr}$), responsible for their red colours.
Moreover, the age distributions of TNG100 galaxies are almost flat at all stellar masses, in particular going towards the high-mass tail, suggesting the presence of similar stellar populations.
In light of that, we do not expect for our sample of old and red galaxies (selected as such as described in \autoref{ssec:ETG_selection}) a significant discrepancy between mass- and luminosity-weighted velocity dispersion profiles.}.
The stellar mass surface density, metallicity and age measurements for MaNGA sources are computed using both \textsc{FIREFLY} and \textsc{Prospector}, while for line-of-sight velocity dispersions we use \textsc{pPXF} (see \autoref{ssec:manga_survey}).
For MaNGA galaxies, the median profiles and related uncertainties in each radial bin are computed as described in \autoref{ssec:median_profiles}, but imposing the condition that the measurements are available for at least $75\%$ of the sample in each stellar mass bin.
\textsc{FIREFLY} and \textsc{Prospector} measurements are derived using stellar population libraries that assume different values of solar metallicity ($\mathrm{Z}_{\odot,\textsc{FIREFLY}}=0.019$ and $\mathrm{Z}_{\odot,\textsc{Prospector}}=0.0142$). For this reason, to homogenise and make comparisons easier, all the observed and simulated stellar metallicity profiles are normalised by their respective values at $\approx7\,\kpc$, which corresponds to the median radius of the the third MaNGA bin.

Since our main goal is to provide a possible evolutionary scenario on how present-day Universe ETGs have been formed throughout the cosmic history, all the stellar properties considered in this work (for both observed and simulated objects) are \emph{at face value}, i.e. as directly derived from pipelines and fitting codes for MaNGA, and from output catalogues for TNG100. Thus, no mock observational data of simulated galaxies have been produced, as instead done for instance by \citet{Nanni2022MNRAS}.

\section{Results}
\label{sec:manga_tng_results}

We now compare the stellar mass surface density, metallicity, age, and line-of-sight velocity dispersion profiles obtained for our observed and simulated samples of ETGs. We focus here on the results corresponding to the stellar mass bins. The results for the number-density-based bins are shown in  \hyperref[app:number_density_cases]{Appendix B}.

\begin{figure*}
    \centering
    \includegraphics[width=.95\textwidth]{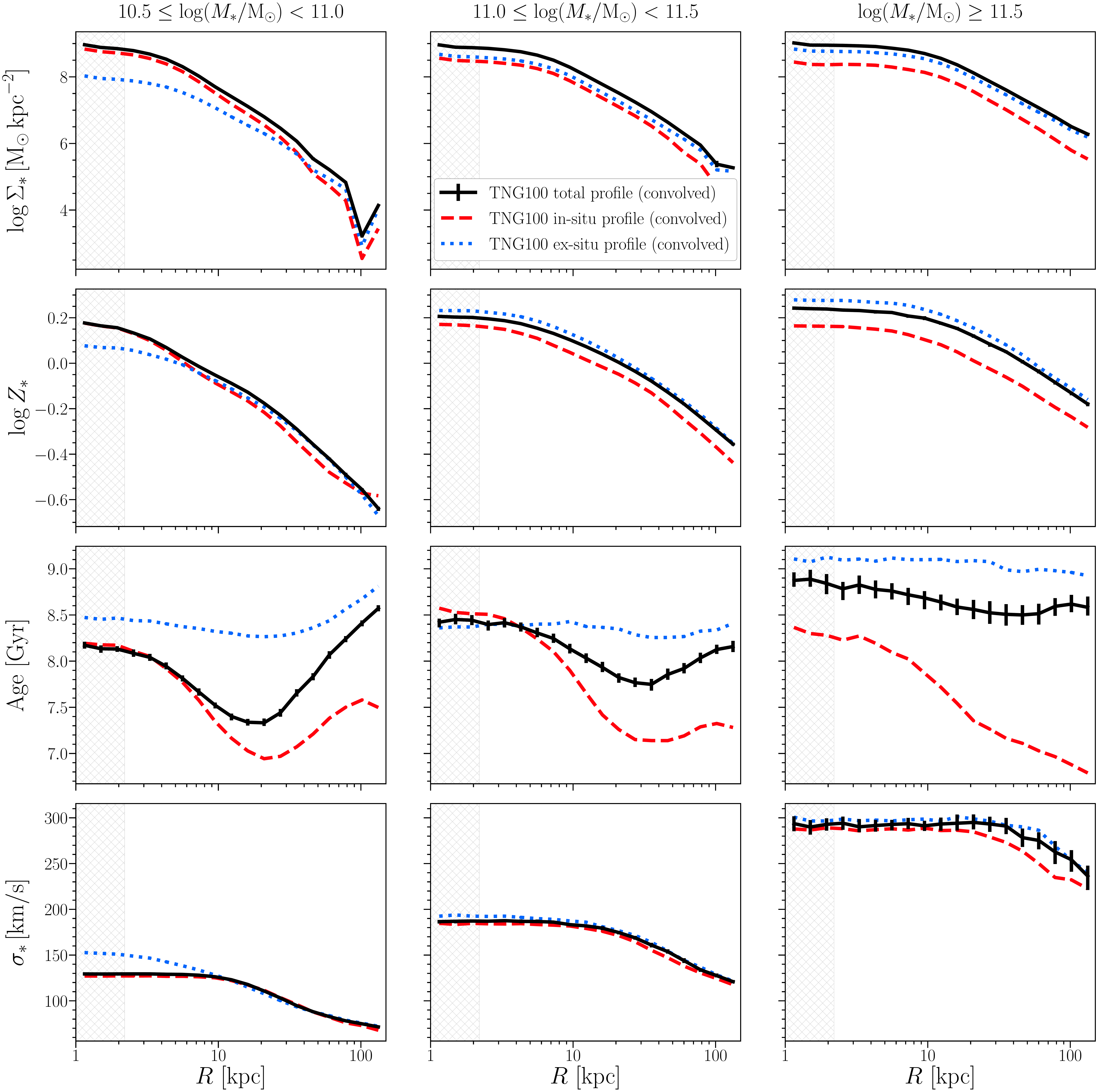}
    \caption{Contribution from in-situ and  ex-situ stellar populations to the radial profiles of TNG100 ETGs. Top to bottom corresponds to profiles of stellar mass surface density, metallicity (not normalised), age, and line-of-sight velocity dispersion. The stellar metallicity, age, and velocity dispersion measurements are mass weighted.
    The black solid, red dashed, and blue dotted curves correspond to the total, in-situ, and  ex-situ stellar populations, respectively. For clarity reasons, here we only show the median profiles convolved with the MaNGA PSF, and we omit the errorbars for the in-situ and ex-situ stellar population profiles. The light grey hatched area ($R\lesssim2.1$ kpc) shows three times the gravitational softening length of the stellar particles in TNG100. \emph{Note}: the range shown along the $y$-axes are different from those of \autoref{fig:profiles_mass_dusty}.}
    \label{fig:profiles_mass_dusty_inex}
\end{figure*}

\subsection{Stellar mass surface density profiles}

\autoref{fig:profiles_mass_dusty} shows the results for the total stellar populations. For TNG100, both the raw profiles and profiles convolved with the MaNGA PSF are displayed. We show profiles both using the median estimates from \textsc{FIREFLY} (yellow dots) and \textsc{Prospector} (green squares). In each stellar mass bin, we find a satisfying agreement at all radii between the two MaNGA measurements, with a small systematic shift to higher values with \textsc{Prospector}. This is consistent with the observed systematic shift in the total mass estimates (for the same galaxy, \textsc{Prospector} infers a stellar mass that, on average, is higher than the estimate obtained by \textsc{FIREFLY} by $\approx0.15{-}0.2\,\mathrm{dex}$). In each stellar mass bin, we find a remarkable agreement between the stellar mass surface densities of MaNGA ETGs and those from TNG100, both in the shape and normalisation of the profiles.

In \autoref{fig:profiles_mass_dusty_inex} we separate the TNG100 profiles into in-situ (red dashed curves) and ex-situ (blue dotted curves) components. In the mass range $10.5<\log(M_*/\msun)<11$, on average, the in-situ stellar component is found to be dominant out to $\approx30\,\kpc$ and ex-situ stars dominate at larger radii. The central mass bin, i.e.\ $11<\log(M_*/\msun)<11.5$, reveals the increasing contribution from ex-situ stars -- the profiles of the two populations contribute in almost equal proportions over the entire radial range. Above $\log(M_*/\msun)\approx11.5$, the ex-situ stars dominate at all radii. 
To summarise, we find that below $\log(M_*/\msun)\approx11$ the most relevant stellar component (out to $\approx30\,\kpc$) is the in-situ population, whereas at higher stellar masses the ex-situ stars become dominant across the entire radial range.

\subsection{Metallicity profiles}

The second row of plots in \autoref{fig:profiles_mass_dusty} displays the metallicity profiles in three stellar mass bins. 
The two MaNGA profiles show similar radial distributions, which differ only beyond $10\,\kpc$ for the most massive systems by a factor $\lesssim0.2\,\mathrm{dex}$ at $\approx 20\,\kpc$\footnote{If the two MaNGA stellar metallicity distributions are not renormalised at their $\sim 7\,\kpc$ values, the \textsc{FIREFLY} profiles, on average, are shifted up from the \textsc{Prospector} profiles by a factor of $\lesssim0.05\,\mathrm{dex}$ at $\log(M_*/\msun)\lesssim{11.5}$, and can differ even of $\lesssim0.16\,\mathrm{dex}$ for galaxies with $\log(M_*/\msun)\gtrsim{11.5}$.}.

Globally, TNG100 profiles reproduce fairly well the shapes of the two MaNGA estimates.
In particular, below $10^{11}\,\msun$, the TNG100 MaNGA-PSF convolved profile reproduces well the median profiles from \textsc{FIREFLY} and \textsc{Prospector}.
Over the interval between $10^{11}\,\msun$ and $10^{11.5}\,\msun$, the median profiles of MaNGA ETGs lie in between the two TNG100 profiles derived from the MaNGA-PSF unconvolved and convolved maps within around $4\,\kpc$, and almost overlap beyond this distance with the two TNG100 profiles. Above $10^{11.5}\,\msun$, the \textsc{FIREFLY} profile is quite well represented even in the outermost regions, while the \textsc{Prospector} distribution tends to assume lower values in metallicity, differing from the TNG100 profiles by a factor of $\lesssim0.1\,\mathrm{dex}$.
We stress here the importance of applying a smoothing using the MaNGA PSF on the original maps of simulated objects. Indeed, at all bins the stellar metallicity profiles from the original maps are steeper than those obtained from the convolved maps, the latter giving values lower by $\approx0.15{-}0.2\,\mathrm{dex}$ in the innermost regions ($R\lesssim4\,\kpc$).

The necessity of renormalising  metallicity measurements to reconcile observations and simulations has been already highlighted by \citet{Nelson2018MNRAS}. The right panel of Figure 2 of \citet{Nelson2018MNRAS} shows the stellar mass-metallicity relations for TNG100 (and TNG300) compared with observed estimates in the present-day Universe from \citet{Gallazzi2005MNRAS,Woo2008MNRAS,Kirby2013ApJ}. Above $\log(\mstar/\msun)>10.5$, simulations and observations almost agree in shape, showing a weak scaling with stellar mass (with a supersolar metallicity normalisation). However, the almost-flat trend of the metallicity as a function of stellar mass in simulations implies a discrepancy of up to $\approx0.5\,\mathrm{dex}$
at $\log(\mstar/\msun)<10.5$ from observed estimates. A possible reason for the origin of such a discrepancy can be found in different methods to derive simulated and observed metallicities. Indeed, when opportune corrections and spectral fitting codes similar to those adopted on observational data are applied to simulated galaxies, the aforementioned discrepancy reduces, making the estimates of TNG100 more consistent with those from observations.


The in-situ and ex-situ PSF-convolved stellar metallicity profiles (second row of plots in \autoref{fig:profiles_mass_dusty_inex}) are shown without adopting any normalisation.
TNG100 galaxies are characterised by ex-situ stars that are more metal rich than the in-situ population:  this metallicity difference increases for increasing stellar mass.
This apparently counter-intuitive finding could be in tension with the expected scenario from downsizing. A possible explanation for the presence of such metal-rich ex-situ stellar populations in these massive systems could be ascribed to the fact that, because of the substantial ex-situ fraction accreted via major mergers (see \autoref{sec:discussion}) across their stellar mass assembly histories, many galaxies that were centrals at a given snapshot became satellites of lightly more massive systems soon thereafter.

\subsection{Age profiles}

\autoref{fig:profiles_mass_dusty} also displays a comparison between the radial distributions of stellar ages. The stellar age profiles derived from \textsc{FIREFLY} and \textsc{Prospector} show a common behaviour in all mass bins: namely, a systematic shift in age is found between the two stellar fitting codes. On average, \textsc{FIREFLY} and \textsc{Prospector} differ in their age estimates by about $1.5{-}2.5\,\mathrm{Gyr}$ (on average, \textsc{FIREFLY}  estimates are $\approx20\%$ younger than those from \textsc{Prospector}).
The systematic difference in age obtained by the two codes might be partially explained in terms of the \emph{age-metallicity degeneracy}: the red colours that characterise old stellar populations can be explained also assuming a higher metallicity, and viceversa \citep{Worthey1994ApJS}. Indeed, for the same sample of ETGs, on average, \textsc{FIREFLY} derives more metal-rich and younger stellar populations compared to \textsc{Prospector}. However, this degeneracy should ideally be reflected in the uncertainty values produced by the stellar population synthesis codes. The fact that the measurements are inconsistent may also suggest that elements of the models used by the two codes, such as the stellar libraries, are themselves inconsistent, and not sufficiently flexible.
A crucial point is that deriving stellar ages for such old systems is not trivial \citep[see][]{Conroy2013ARA&A}. Indeed, stellar age grids for these models are sparse at these ages because they tend to be log-spaced. When building non-parametric SFHs, these codes interpolate over the ages sampled by the stellar libraries and isochrones. 
The large gap between the two stellar age profiles displayed in \autoref{fig:profiles_mass_dusty} can be taken as a measure of the systematic uncertainty on the age of the observed galaxies in our sample. It is clear that these age profiles have little constraining power on theoretical models: the TNG100 profiles lie in between \textsc{FIREFLY} and \textsc{Prospector}, reproducing only the quasi-flat distributions of observed data. The analysis of the radial distributions of age for the in-situ and ex-situ stellar populations in the simulated ETGs (see \autoref{fig:profiles_mass_dusty_inex}) shows that, below $\log(\mstar/\msun)\approx11$, the ex-situ component is older (up to $+1.5\,\mathrm{Gyr}$) than the in-situ component over the entire radial range, whereas the inner regions ($R\lesssim6\,\kpc$) of galaxies with $11\lesssim\log(\mstar/\msun)\lesssim11.5$ are composed of in-situ and ex-situ stars with similar ages.
Above $\log(\mstar/\msun)\approx11.5$, ex-situ stars are found to be older (up to $+2.5\,\mathrm{Gyr}$) than the in-situ population at all radii.

\subsection{Velocity dispersion profiles}

The bottom panels in \autoref{fig:profiles_mass_dusty} compares the radial profiles of line-of-sight stellar velocity dispersions for simulated and observed ETGs. Here, the MaNGA values are derived using the \textsc{pPXF} code.
Below  $\log(\mstar/\msun)\approx11.5$, we generally find a good first-order agreement in normalisation between MaNGA and TNG100, but the MaNGA profiles are steeper than those of TNG100. More quantitatively, the difference between the two median profiles can be at most of around $30\,\kms$.
Instead, above $\log(\mstar/\msun)\approx11.5$, the velocity dispersion profiles of both MaNGA and TNG100 galaxies are almost flat out to $R\approx40\,\kpc$. However,  we find an essentially radius-independent difference between the two profiles, with velocity dispersions for the simulated ETGs generally higher by $30{-}40\,\kms$ (this will be discussed further in \autoref{ssec:central_satellite}).
We underline that the effect on the observed velocity dispersion $\sigma_\mathrm{*}$ for MaNGA galaxies is the result of two main contributions, i.e. the instrumental dispersion $\sigma_\mathrm{inst}$ and the intrinsic stellar velocity dispersion $\sigma_\mathrm{*,int}$:
\begin{equation}
    \sigma_\mathrm{*}^2=\sigma_\mathrm{inst}^2+\sigma_\mathrm{*,int}^2.
\end{equation}
In fact, over the rest-frame optical range $0.36\lesssim\lambda/\mathrm{\mu m}\lesssim1.03$, the spectral resolution is $R\sim2000$ and the $1\sigma$ dispersion of the instrumental spectral line-spread function is about $70\,\mathrm{km/s}$ \citep[see][]{Westfall2019AJ,Law2021AJ}. As reported in \citet{Westfall2019AJ}, for $\sigma_*\gtrsim100\,\mathrm{km/s}$, the uncertainties on velocities can be approximated as $\delta(\Delta v)\approx \langle\sigma_*\rangle/(S/N)_g$, where $\langle\sigma_*\rangle$ is the mean velocity dispersion and $(S/N)_g$ is the $g$-band signal-to-noise ratio. At $(S/N)_g=10$, the typical uncertainties on velocities are around $10\%$ of $\sigma_*$. For $\sigma_*\gtrsim100\,\mathrm{km/s}$, the uncertainties on velocity dispersions are slightly larger the those of velocities, but can be roughly approximated by a single proportionality constant.
Given that, our estimates of velocity dispersion profiles are quite robust, because, on average, these measurements are greater than $100\,\mathrm{km/s}$, with the only exception of the last radial bin of the profile for galaxies with $10.5\leq\log(M_*/\mathrm{M_\odot})<11$, that is around $90\,\mathrm{km/s}$.

The line-of-sight stellar velocity dispersion profiles for the in-situ and ex-situ components (bottom panels in \autoref{fig:profiles_mass_dusty_inex}) almost coincide in the 
intermediate and high-mass bins, while in the low-mass bin the velocity dispersion is higher in the centre for the in-situ component.
As it is well known, in the same gravitational potential, the line-of-sight stellar velocity dispersion profile of a given component depends on both its intrinsic velocity distribution and its density distribution: in particular, for given velocity distribution, the steeper the density profile, the lower the velocity dispersion (see \citealt{Nipoti2021MNRAS}).
In the low-mass bin, the higher central velocity dispersion of the in-situ component can be qualitatively explained by its shallower surface density profile (see top-left panel in \autoref{fig:profiles_mass_dusty_inex}).


\subsection{Central versus satellite galaxies}\label{ssec:central_satellite}

We now consider the radial profiles separately for central and satellite galaxies. 
Halos and subhalos in IllustrisTNG are detected by \textsc{subfind}, the subhalo finder code developed by \citet{Springel2001MNRAS}. Specifically, an IllustrisTNG subhalo is classified as \emph{central} (flag {\ttfamily is\_primary==1}) if it is 
the subhalo with the deepest potential well among those belonging to the same friends-of-friends (FoF) halo. Otherwise, subhalos are classified as \emph{satellites}. To separate MaNGA ETGs into centrals and satellites we rely on the classification provided by \citet{Yang2007ApJ}, obtained for a sample of more than 300000 galaxies from SDSS DR4 \citep{Adelman-McCarthy2006ApJS}. 

Though they are not shown here, we do find that the MaNGA and TNG100 stellar mass surface density profiles, as well as those for the stellar metallicity and age, are in excellent agreement between central and satellite ETGs. This lack of differences in the radial profiles between these two populations is consistent with previous outcomes in literature, as for instance shown by \citet[][]{Santucci2020ApJ} for stellar age and metallicity properties.

Centrals and satellites display similar trends also for the in-situ and ex-situ components over the entire mass range considered. We might expect central galaxies to exhibit larger ex-situ components compared to satellites. However, a  possible explanation for the observed similarity could be that most satellites in the considered mass range were recently accreted onto the main halo and acquired a significant fraction of their ex-situ component when they were centrals of other halos.

\begin{figure*}
    \begin{subfigure}{1\textwidth}
    \centering
    \includegraphics[width=.9\textwidth]{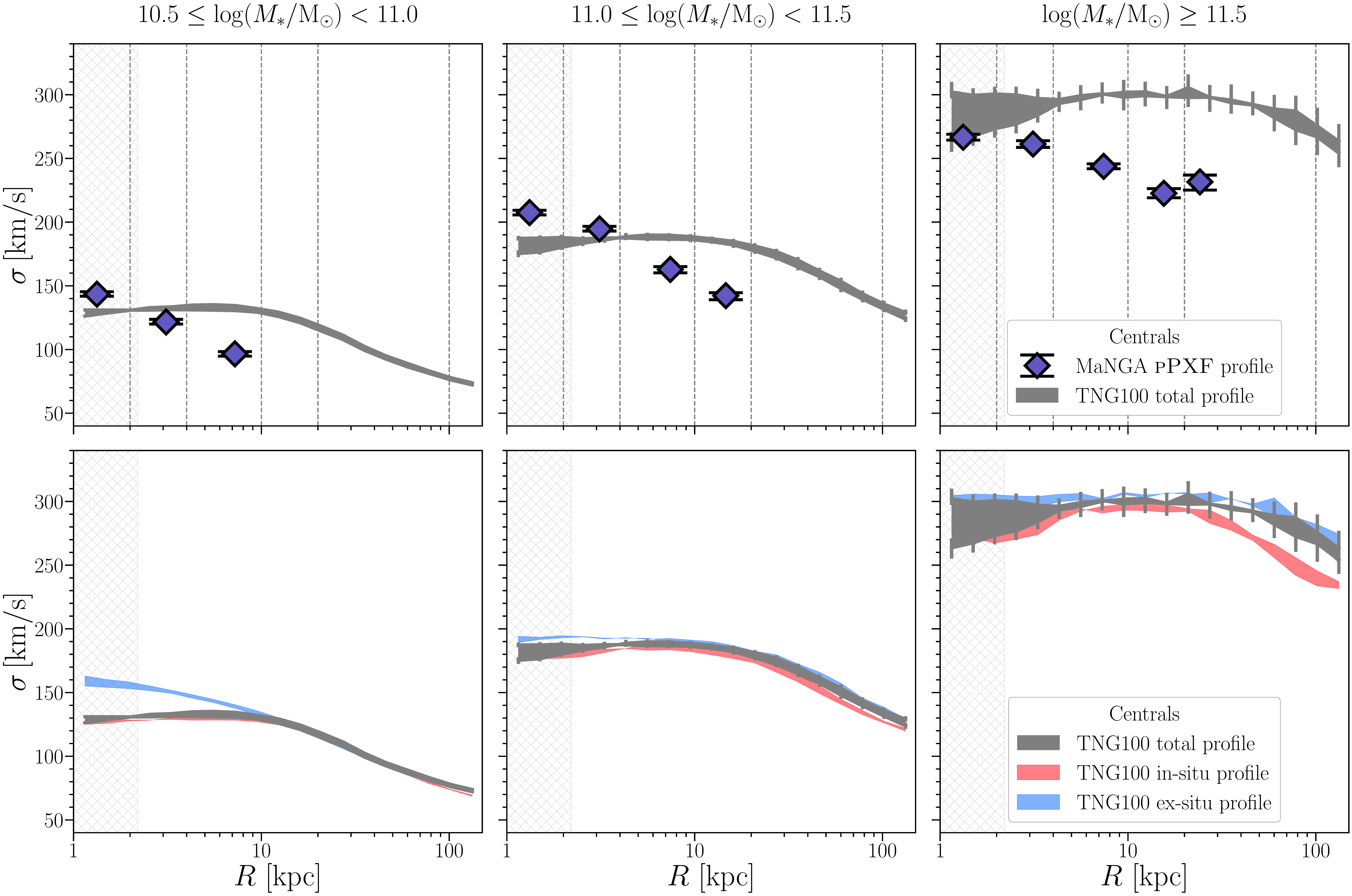}\\
    \end{subfigure}\hfill
    \begin{subfigure}{1\textwidth}
    \centering
    \includegraphics[width=.9\textwidth]{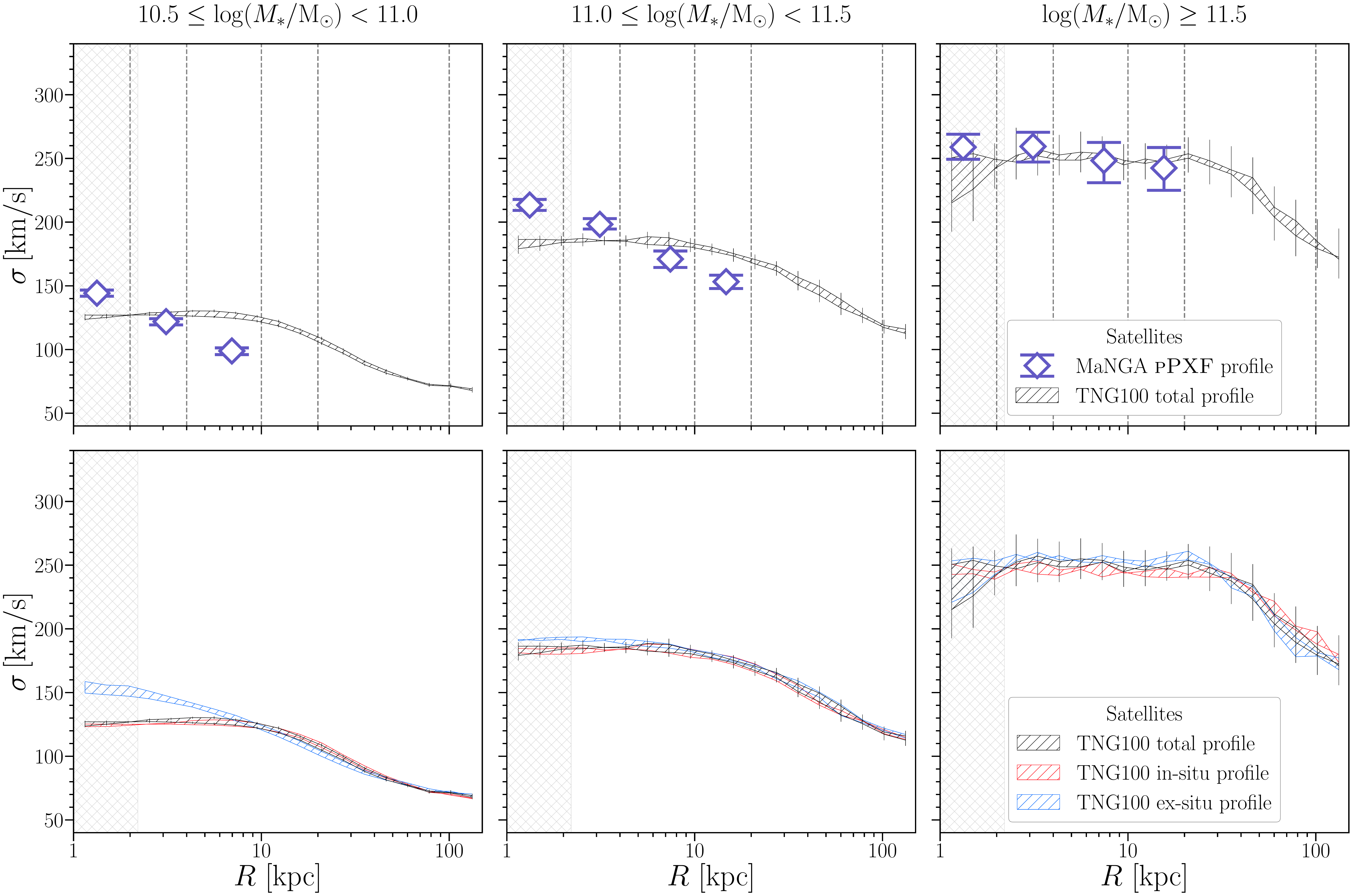} 
    \end{subfigure}
    \caption{Line-of-sight stellar velocity dispersion radial profiles for centrals (upper rows) and satellites (lower rows). The violet-filled black and white-filled violet diamonds represent the median estimates of the mass-weighted velocity dispersion profiles for MaNGA central and satellite ETGs from \textsc{pPXF} code, respectively. The vertical grey dashed lines delimit the 5 radial intervals within stellar velocity dispersion is computed for MaNGA ETGs.
    The shaded and hatched areas represent the regions delimited by the median values of the mass-weighted velocity dispersion profiles obtained from the original TNG100 maps and the maps convolved with the MaNGA PSF for centrals and satellites, respectively. Black, red and blue colours correspond to the total, in-situ, and ex-situ stellar populations. For clarity reasons, we show only the errobars for the MaNGA and TNG100 total stellar population profiles.
    The light grey hatched area ($R\lesssim2.1$ kpc) shows three times the gravitational softening length of the stellar particles in TNG100.}
    \label{fig:profiles_centrals_satellites_vdisp}
\end{figure*}


\autoref{fig:profiles_centrals_satellites_vdisp} compares the line-of-sight stellar velocity dispersion profiles for centrals and satellites separately and displays a key result in this paper. Whereas for MaNGA we find that massive centrals and satellites show similar velocity dispersion profiles, in contrast, for the most massive bin we find that TNG100 predicts a $\approx50\,\kms$ offset between centrals and satellites.
The amount of DM in central simulated galaxies could provide an explanation for the significant difference in velocity dispersion between TNG100 centrals and satellites. As discussed in \citet{Lovell2018MNRAS},  TNG100 predicts an important enhancement of the DM content in the inner regions of subhalos. Hence, this high fraction of DM, that dominates galaxies at $z\approx0$, may be the responsible of this high velocity dispersion especially for the most massive central galaxies.
Our observations appear to exclude a difference in the velocity dispersion profile between centrals and satellites. However, determining whether a galaxy is a central or a satellite is notoriously difficult, and misclassifications in the \citet[][]{Yang2007ApJ} catalogue may erase the observational signal. This possibility warrants further investigation before firm conclusions can be drawn.


\subsection{Robustness of Results}

We now present a discussion on the robustness of the results.

\begin{itemize}
\item \emph{Sample matching method} $-$ MaNGA galaxies can be matched to simulated galaxies according to stellar mass or number density. We have tried both and found only minor differences. These are discussed further in  \hyperref[app:number_density_cases]{Appendix B}.\\

\item \emph{Different definitions of stellar masses} $-$
Given the wide variety of possible systematic effects on the definition of stellar masses, in Chapter 3 of \citetalias{Cannarozzo2021} the same analysis was performed by testing also the SerExp Dust-free stellar masses \citep[from][]{Mendel2014ApJS}, the Sérsic and Petrosian fit estimates from the original NSA catalogue, and the masses defined as the sum of the masses included in the 5 concentric annuli used to derive the profiles of stellar properties  from \textsc{FIREFLY} and \textsc{Prospector} for MaNGA ETGs, while for TNG100 galaxies we considered also the stellar masses within a projected aperture of $30\,\kpc$. These tests do not reveal significant differences from the analysis presented in this work.\\

\item \emph{Different definitions of ETGs} $-$ Also the ETG selection is a factor that could affect the results of this study. In Appendix A of \citet{Diemer2019MNRAS}, the authors compare the ETG fractions derived from diverse selection methods in IllustrisTNG with that from the observed compilation of \citet{Calette2018RMxAA}. In particular, the authors measure the ETG fractions adopting classifications based on the concentration of the 3D stellar mass density profiles, $C_{82}$, defined as $5\times\log(r_{80}/r_{20})$, with $r_{80}$ and $r_{20}$ as the radii including the $80\%$ and $20\%$ of the total stellar mass, on $(g-r)$ colours, on spheroid-to-total ratios $S/T$, and on 
the fraction of kinetic energy that is in rotation $\kappa_\mathrm{rot}$. They found that the best indicator of galaxy morphology able to better reproduce the to reproduce the ETG fraction from \citet{Calette2018RMxAA} is $C_{82}$. Instead, the $(g-r)$ classification implies an excess of ETGs, and that the colours correlate weakly with structural parameters \citep[as illustrated also in ][]{RodriguezGomez2019MNRAS}.
\citet{Tacchella2019MNRAS} studied the connection between the star formation activity and morphology of central galaxies in IllustrisTNG, adopting as morphological indicators the parameters $C_{82}$ and $S/T$. They found that the $S/T$ parameter strongly correlates with $(g-r)$ colours: $S/T$ is higher for redder colours and higher stellar masses (while, at fixed mass, $C_{82}$ is found to be weakly dependent on colour).
Our choice of adopting a simple selection on colours is driven by the fact that mock colours in IllustrisTNG are generated consistently with observations (we remind that these mock colours are obtained following the observational prescriptions described in section 3 of \citealt{Nelson2018MNRAS}).
As done for the different stellar mass definitions, in Chapter 3 of \citetalias{Cannarozzo2021} we adopted another ETG selection for both the MaNGA and TNG100 samples, including only those objects with SFRs below 1 dex from the star-forming main sequence of galaxies (one of the methods to select passive systems presented in \citealt{Donnari2019MNRAS}). Based on this extensive exploration, even the results here presented are robust and independent of the specific definitions of ETGs for both the observed and simulated sources.

As mentioned above, the selection of ETGs may involve different criteria, each of which may introduce some selection biases \citep[see also][]{Moresco2013A&A}. Usually, ETGs are characterised by either an elliptical (E) or a lenticular (S0) morphology. One of the historical criteria for morphological selection of galaxies is that based on the T-$\mathrm{Type}$ \citep{deVaucouleurs1959HDP}. According to the T-$\mathrm{Type}$-based classification, E/S0 galaxies have values between $-6$ and $-1$, while the various types of spiral galaxies range between 0 and 9.
In light of that, for MaNGA objects we also verified the impact of the adopted selection based on colours (see \autoref{ssec:ETG_selection}), checking the morphological type assigned by the \textsc{MaNGA Morphology Deep Learning DR15 catalog}\footnote{Available at \url{https://www.sdss.org/dr15/data_access/value-added-catalogs/?vac_id=manga-morphology-deep-learning-dr15-catalogue}}. This catalogue, presented in \citet{Fischer2019MNRAS}, is built by exploiting the Deep Learning method for identifying the morphology of galaxies as described in \citet{DominguezSanchez2020MNRAS} for all the objects of MaNGA DR15.
In our Red Galaxy sample, $\approx66\%$ of the total effectively shows a clear morphology compatible with an E/S0 type, of which $\approx33\%$ are classified as lenticular galaxies.
\end{itemize}

\section{Discussion}
\label{sec:discussion}
In this section we discuss our results and compare our findings with previous works. 

\subsection{The role of mergers in TNG100 galaxies}
\label{ssec:role_of_mergers}
\autoref{fig:profiles_mass_dusty_inex} shows that the shapes of the in-situ and ex-situ stellar mass surface density radial profiles from TNG100 ETGs are stellar-mass dependent and that the ex-situ component dominates the total profile at all radii above $\log(M_*/\msun)\approx11$. This agrees with previous results from \citet{Pillepich2018MNRAS}, \citet{Tacchella2019MNRAS}, and \citet{Pulsoni2021A&A}, who find that the stellar mass assembly history of very massive galaxies is driven by major mergers. Indeed, major mergers not only allow ex-situ stellar populations to settle even in the innermost regions of galaxies, but also homogeneously mix the two stellar components at all radii, causing the formation of stellar mass surface density profiles which are similar in shape and differ only in their normalisation.
However, the results of IllustrisTNG disagree with some previous works finding that the growth of massive ETGs is primarily driven by minor mergers. \citep[e.g., ][]{Naab2009ApJl,Oser2010ApJ,Hilz2013MNRAS}. For example, as argued in \citet[][]{Genel2008ApJ}, and in \citet{KhochfarSilk2009MNRAS}, massive ($\gtrsim10^{11}\,\msun$) DM halos undergo typically no more than one major merger in the redshift range $0\lesssim z\lesssim2$. There aro also some observational works supporting the idea that massive ETGs may experience few major mergers \citep[e.g.,][]{Bell2006ApJ,McIntosh2008MNRAS}, instead undergoing a high number of minor mergers \citep[e.g.,][]{Bundy2009ApJ}.  On the contrary, other studies lends support to the idea in which the role of major mergers may be more relevant, estimating relatively high mass-weighted merger ratios. For example, \citet[][]{SonnenfeldNipotiTreu2017MNRAS} infer for galaxies of $\log(M_*/\msun)\approx11$ a mass-weighted merger ratio greater than 0.4. 


\begin{figure*}
    \centering
    \includegraphics[width=1\textwidth]{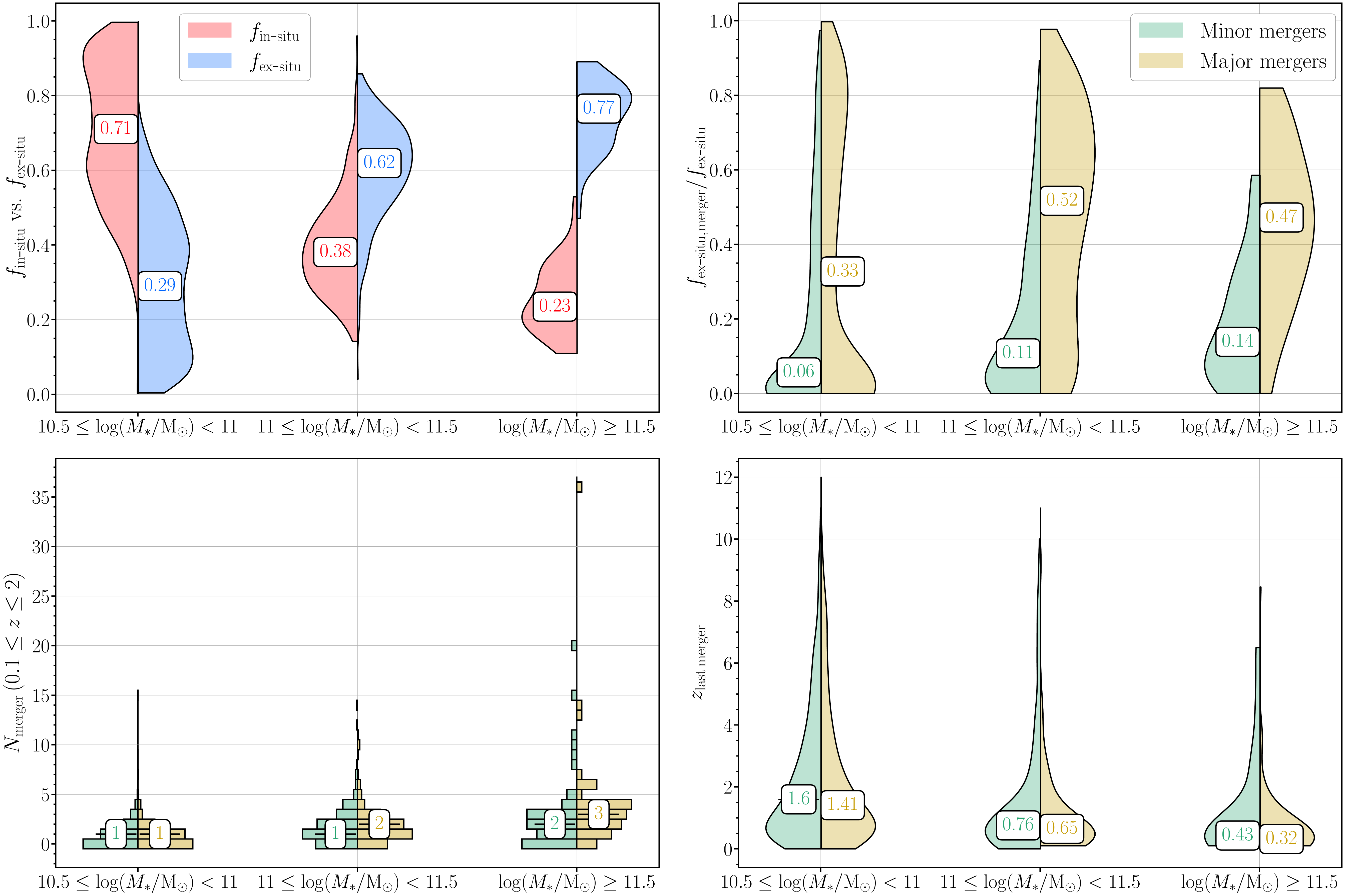}
\caption{Data distributions of eight properties related to the merger histories 
of TNG100 ETGs. Top-left panel: total fractions of the in-situ (red distributions) and the ex-situ (blue distributions) stellar components. Top-right panel: fractions of the ex-situ stellar component from minor (green) and major (ocre) mergers normalised to the total ex-situ stellar fraction. Bottom-left panel: number of minor (green) and major (ocre) mergers across the redshift range $0.1\leq z\leq2$. Bottom-right panel: redshifts of the last minor (green) and major (ocre) mergers. All distributions are shown for the same three stellar mass bins as in \autoref{fig:profiles_mass_dusty} and in \autoref{fig:profiles_mass_dusty_inex}.
Except for the number of minor and major mergers which are discrete values, the other properties are displayed as violin plots.
For each property, the median value of the corresponding distribution is reported.}
    \label{fig:mergers_tng}
\end{figure*}

\autoref{fig:mergers_tng} displays how minor and major mergers contribute to the the total mass of TNG100 ETGs as a function of redshift and galaxy mass.
For each stellar mass bin, the data distributions are presented as \emph{violin plots} (except for the number of mergers), which displays the probability density of the data smoothed by a kernel density estimator.
The shape of each violin plot represents the frequency of data, so that the larger 
the violin's body, the higher the density of data at a given $y$-axis value.
Specifically, we make use of 100 data points to evaluate each Gaussian kernel density estimation.
The complementary behaviour of the in-situ and ex-situ stellar mass fractions in the top-left panel of \autoref{fig:mergers_tng} confirms the rising importance of accreted stars in higher-mass ETGs. In particular, the median of the ex-situ stellar fraction grows from  $\approx29\%$ for galaxies with $10.5\leq\log(M_*/\mathrm{M}_\odot)<11$, up to $\approx 77\%$ for the most massive systems.
Following \citet{RodriguezGomez2015MNRAS,RodriguezGomez2016MNRAS}, $\mu_*$ is defined as the \emph{stellar mass ratio} between the two progenitors of a given galaxy. A \emph{major merger} is then defined by a $\mu_*>1/4$, while a \emph{minor merger} is defined by $1/10<\mu_*<1/4$. 
However, the fraction of accreted stars from other galaxies is not only due to major and minor mergers. It also includes stars from the so-called \emph{very minor mergers}, i.e. with $\mu_*<1/10$, as well as tidally stripped stars from surviving galaxies.
As illustrated in the top-right panel of \autoref{fig:mergers_tng}, the analysis of the ex-situ fraction accreted via minor mergers relative to the total ex-situ fraction, reveals that, on average, around $6\%$ of stars are accreted via minor mergers in ETGs with $M_*<10^{11}\,\msun$, reaching $\approx14\%$ in the most massive galaxies. By isolating the role of major mergers, their contribution presents broad distributions, with median values of around $50\%$ (relative to the whole ex-situ stellar fraction) for ETGs with $\log(M_*/\mathrm{M}_\odot)>11$.
The analysis of the relative contributions deriving from both minor and major mergers suggests that, on average, below $10^{11}\,\msun$, TNG100 galaxies accrete the majority of their ex-situ stellar population through very minor mergers and by stripping stars from surviving objects. Indeed, for these systems $(f_\mathrm{ex\text{-}situ,\, minor\,merger}+f_\mathrm{ex\text{-}situ,\, major\,merger})/f_\mathrm{ex\text{-}situ}\approx39\%$. Instead, above $10^{11}\,\msun$, more than $60\%$ of the ex-situ component comes from minor and major mergers, with a larger contribution from major mergers.
Considering the distributions of the number of minor and major mergers in TNG100 ETGs for galaxies below $10^{11}\,\msun$ over the redshift range $0.1\leq z\leq2$, below $10^{11}\,\msun$,  ETGs usually undergo at most one minor and/or major mergers, while above $10^{11}\,\msun$ the distributions are slightly wider, confirming the important role of major mergers in shaping massive galaxies. 
Finally,  TNG100 ETGs experience, on average, their last major mergers slightly more recently than their last minor mergers.
However, we stress here that by minor mergers we refer to systems with $1/10<\mu_*<1/4$, which excludes the very minor mergers (i.e. $\mu_*<1/10$). In addition, we draw attention to the fact that, at a given stellar mass, both the distributions and their median values of the last minor and major mergers are similar, implying that the differences are fairly small. The difference between the median values for the same subsample are lower than $1\,\mathrm{Gyr}$.

\subsection{Comparison with Recent Works}

In \autoref{sec:manga_tng_results}, we discussed the comparison of the circularised radial distributions of the stellar mass surface density, metallicity, age, and velocity dispersion between in MaNGA and TNG100 ETGs.
In this section, we compare our results with other works from the literature.

Using a sample of 366 ETGs with masses in the range $9.9<\log(\mstar/\msun)<10.8$ selected via the the Galaxy Zoo morphological classification \citep{Lintott2011MNRAS,Willett2013MNRAS}, plus visual inspection, \citet{Parikh2018MNRAS,Parikh2019MNRAS} analysed the radial gradients of stellar age and metallicity out to one effective radius. If we consider in our sample only galaxies with stellar mass lower than $10^{10.8}\,\msun$ and we rescale our profiles in units of $\reff$ (the median $\reff$ for our MaNGA ETGs with stellar mass lower than $10^{10.8}$ is $\lesssim3\,\kpc$) to directly compare the results with those from \citet{Parikh2018MNRAS,Parikh2019MNRAS}, we find a satisfying consistency with their stellar age and metallicity radial distributions.

\citet{Bernardi2019MNRAS} show stellar population gradients for a sample of MaNGA DR15 ETGs subdivided into slow and fast rotators.
These ETGs are identified as in \citet{DominguezSanchez2020MNRAS}, i.e.\ applying a morphological classification based on T-$\mathrm{Type}\leq0$ (see also \autoref{ssec:Stellar_mass_estimates}), considering both pure ellipticals and lenticulars.
The stellar age and metallicity gradients measured by \citet{Bernardi2019MNRAS} out to $1\,\reff$ are qualitatively compatible with our estimates. One of the most relevant outcomes of \citet{Bernardi2019MNRAS} is that slow rotators dominate above $\log(M_*/\msun)\approx11.5$: at this stellar mass, where also the size-mass relation slope changes \citep[see][]{Bernardi2011MNRAS}, the majority of these ellipticals are central galaxies\footnote{\citet{Bernardi2019MNRAS} make use of the \citet{Yang2007ApJ} environmental catalogue used also in this work. \label{footnote}}.
As an extension of the \citet{Bernardi2019MNRAS} work, \citet{DominguezSanchez2020MNRAS} focus on stellar properties of S0 lenticular galaxies, highlighting a bimodality in this galaxy population that depends on stellar mass. Above $\log(M_*/\msun)\approx10.5$, indeed, these galaxies are characterised by stronger age and velocity dispersion gradients, with, instead, negligible gradients in metallicity.

Recently, \citet{BarreraBallesteros2022arXiv} analysed the entire set of around 10000 galaxies from the MaNGA survey, presenting the radial distributions of several physical properties, selecting in particular a subset of about 1400 sources with optimal spatial coverage, for which the authors studied the impact of a selection based on either stellar mass or morphology. Among the properties derived through the the newest \textsc{pyPipe3D} pipeline \citep{Sanchez2022arXiv}, the authors measured the radial distributions of stellar mass surface density, luminosity-weighted stellar metallicity and age, and velocity dispersion. The negative gradients found from \citet{BarreraBallesteros2022arXiv} for the stellar mass surface density and metallicity in elliptical and lenticular galaxies are qualitatively in agreement with the measurements obtained in this manuscript, reflecting also the increasing of the normalisations of both profiles as the stellar mass goes up. Even the velocity dispersion profiles agree with those computed by \textsc{pPXF}, finding decreasing distributions towards the outer regions, and central measurements of $\sim 150\,\kms$ for E/S0 galaxies with $10.5<\log(M_*/\msun)<11$, and of $\sim 250\,\kms$ for objects with $\log(M_*/\msun)>11$, similarly to those presented here.
Regarding stellar ages, the profiles in \citet{BarreraBallesteros2022arXiv} show values slightly more consistent with those derived by \textsc{Prospector} (i.e., $\gtrsim9\,\mathrm{Gyr}$) presented in this paper, though their radial distributions tend to reduce at large distances, implying the presence of younger stellar populations in the outer regions of galaxies.
Also \citet{Oyarzun2022ApJ} present the radial profiles for stellar mass surface density from \textsc{Prospector}, as well as for element abundances (i.e. $[\mathrm{Fe/H}]$ and $[\mathrm{Mg/Fe}]$) and ages from  \textsc{alf} (\citealt{Conroy2018ApJ,ConroyVanDokkum2012ApJ}; see also \citealt{Conroy2014ApJ,Choi2014ApJ}) for a subset of about 2200 passive centrals from MaNGA to understand the impact of stellar and halo masses in assembling these systems. 
The radial distributions of stellar mass surface density and $[\mathrm{Fe/H}]$ (that we can consider as a measurement of the stellar metallicity) are consistent with those shown in \autoref{fig:profiles_mass_dusty}. The ages provided by \textsc{alf}, on average, lie in between our measurements from \textsc{FIREFLY} and \textsc{Prospector}, showing also the presence of younger stellar populations in the inner regions.

Using the results from the Illustris simulation, \citet{Cook2016ApJ} investigated the stellar population gradients for a sample of more than 500 ETGs with $10\leq\log(\mstar/\msun)\leq12$. The stellar surface brightness, metallicity, and age gradients are overall in agreement with observables. The gradients are subdivided into three intervals: the inner galaxy ($0.1{-}1\,\reff$), the outer galaxy ($1{-}2\,\reff$), and the stellar halo ($2{-}4\,\reff$). Except for the age gradients which are found to be not so informative about the accretion histories of galaxies, both the surface-brightness and metallicity profiles show that, at fixed stellar mass, the ex-situ stellar component produce flatter profiles. In particular, as the stellar mas increases, the higher the accreted star fraction, the flatter the profiles.
Though the flattening at large radii of the stellar mass surface density and metallicity profiles is not apparent in  \autoref{fig:profiles_mass_dusty}, because of the adopted logarithmic scale, we verified that our profiles are quantitatively consistent with those of \citet{Cook2016ApJ} when a linear scale in units of $\reff$ is adopted.
Albeit a qualitative agreement in the behaviour of radial profiles is found using Illustris and IllustrisTNG, the fraction of accreted stars as well as the role of mergers in shaping galaxies are significantly different. As highlighted in Figure 10 of \citet{Tacchella2019MNRAS}, though in both simulations the fraction of ex-situ stars at $z=0$ rapidly increases above $10^{10.5}\,\msun$, TNG100 predicts a fraction that, on average, is higher by a factor of $\approx30\%$ with respect to the total stellar amount. Moreover, while the main channel for the stellar accretion in Illustris is via minor mergers, as shown in the top-right panel of \autoref{fig:mergers_tng}, TNG100 predicts a more relevant role of major mergers. These two substantial differences between Illustris and IllustrisTNG are primarily due to the diverse feedback model implemented and the consequent stellar mass functions.
In a more recent work, \citet{Pulsoni2020A&A} studied the photometric and kinematic properties out to $15\,\reff$ of ETGs stellar halos for 1114 objects in TNG100 (together with other 80 sources in TNG50), with stellar masses $10.3<\log(M_*/\msun)<12$ and selected in $(g-r)$ colours (similarly with the selection adopted in this paper) and in the angular momentum--ellipticity plane.
Analogously to our findings, the high-mass tail of ETGs are everywhere dominated by the accreted stellar component, mainly acquired through major mergers. In addition, IllustrisTNG ETGs are compared with some observational surveys, including MaNGA galaxies. Looking at the distribution of galaxies in the angular momentum--ellipticity plane within $1\,\reff$, a percentage of the IllustrisTNG galaxies lie in a region where no observed ETGs are found: these are basically elongated, triaxial systems. However, when simulated galaxies with an intermediate-to-major axis ratio $<0.6$ at $1\,\reff$ are removed - the centrally elongated objects -, these ETGs reflect the location in the plane of the observed counterpart, except for a region where a large fraction of MaNGA S0 have a high angular momentum.

Overall, the results presented in this manuscript are generally in agreement with previous works in the literature. In particular, the distributions of the stellar properties for the observed galaxies analysed in this work confirm the common trend of negative gradients for stellar mass surface density, metallicity, and velocity dispersion, whose normalisations increase as the stellar mass increases.
The radial distributions of stellar age, however, are not always in agreement with those presented in previous works. The differences, whether in normalisation or in shape, may depend on multiple factors, such as the considered samples or the stellar fitting codes and libraries used to estimate the age of stellar populations.
The scope of this work is to outline a possible scenario for the merger-driven evolution of observed ETGs in the present-day Universe. The scenario predicted by IllustrisTNG in which major mergers may be crucial in shaping massive galaxies at $z\approx0$ is somewhat in contrast with some previous theoretical findings \citep[e.g.,][]{Naab2009ApJl,Oser2010ApJ,Hilz2013MNRAS} which, conversely, back an evolution mainly driven by a high number of minor mergers.

\section{Summary and Conclusions}
\label{sec:conclusion_manga_tng}

In this paper we studied the radial profiles of stellar mass surface density, metallicity, age, and line-of-sight velocity dispersion in massive ($M_*\geq10^{10.5}\,\msun$) ETGs, selected in colours with $(g-r)>0.6$, comparing observed galaxies from the MaNGA DR15 survey with simulated galaxies from TNG100 of the IllustrisTNG magneto-hydrodynamical cosmological simulation suite. For both galaxy samples, the stellar property profiles have been obtained building concentric elliptical annuli on the bi-dimensional projected maps of each source, disentangling in TNG100 ETGs the in-situ and ex-situ stellar populations, and both accounting or not accounting for the effects of convolving the maps by MaNGA PSF. All the presented stellar population properties are at face value, i.e. the measurements are directly obtained from pipelines and stellar fitting codes for MaNGA, and from the simulation for TNG100 ETGs.

Our main results are the following.
\begin{itemize}
\item We find a satisfying agreement between observations and simulations in the stacked radial profiles of the stellar mass surface density of massive ETGs, {both in shape and normalisation}. This agreement is observed at all radii and at all stellar mass bins, and is independent of ETG and stellar mass definitions.
\item Overall, TNG100 ETGs have, on average, stellar metallicity and velocity dispersion profiles reasonably similar to those observed in  MaNGA ETGs.
Concerning metallicity, the shape of MaNGA profiles is well reproduced by TNG100 galaxies, though in some cases the observed and simulated profiles differ in normalisation (around $0.15\,\mathrm{dex}$ in the outermost parts of ETGs with $\mstar\gtrsim10^{11.5}\,\msun$).
For galaxies $\mstar<10^{11.5}\,\msun$, we find a decent agreement for the radial distributions of velocity dispersion between simulated and observed ETGs, the latter showing steeper profiles that differ at most by $\approx\,30\,\mathrm{km\,s^{-1}}$ from the simulated ETG distributions. Only the very massive  ($\mstar\gtrsim10^{11.5}\,\msun$) systems of TNG100 tend to have,  over the entire explored radial range, higher velocity dispersion than the corresponding observed system, by up to $\approx 50\,\kms$.

\item The ages of the stellar populations of observed ETGs are highly uncertain, and significantly different age estimates are obtained using different codes (\textsc{FIREFLY} and \textsc{Prospector}, which differ of about $2{-}2.5\,\mathrm{Gyr}$ at all radii and at all stellar mass bins). The age profiles of stellar populations in TNG100 are found to lie in between the profiles estimated for the corresponding observed ETGs with \textsc{FIREFLY} and \textsc{Prospector}.

\item By separating central and satellite galaxies for both TNG100 and MaNGA samples, we find that there are not relevant differences in all the profiles between the two galaxy populations, except for the velocity dispersion profiles of massive systems ($\mstar>10^{11.5}\,\msun$, see \autoref{fig:profiles_centrals_satellites_vdisp}). Indeed, while TNG100 and MaNGA satellites have similar velocity dispersion profiles, central simulated galaxies tend to have velocity dispersion at all radii higher than observed ETGs ($\approx 50\,\kms$).
\item The behaviour of the in-situ and ex-situ surface density profiles identifies two different scenarios for the merger-driven history of these objects, corroborating previous outcomes in the literature, such as from \citet{Pillepich2018MNRAS} and \citet{Tacchella2019MNRAS}: galaxies with $\mstar\lesssim10^{11}\,\msun$ are mainly dominated by the in-situ stellar populations out to $\approx30\,\kpc$; instead, in ETGs with $\mstar\gtrsim10^{11}\,\msun$, the contribution of the ex-situ stars is at least as important as that of the in-situ component, and even totally dominating for very massive ETGs ($\mstar\gtrsim10^{11.5}\,\msun$).
\item The similar shapes found for the radial distributions of the stellar mass surface densities for both in-situ and ex-situ stars (see \autoref{fig:profiles_mass_dusty_inex}) as well as the detailed analysis of the merger history (see \autoref{fig:mergers_tng}) in simulated ETGs reveal that especially galaxies with $\mstar\gtrsim10^{11}\,\msun$ experienced across cosmic time an evolution mainly driven by major mergers. Indeed, major mergers allow both to explain the presence of a significant percentage of ex-situ stars that are able to penetrate even in the innermost parts of galaxies, and also that  the two stellar components are well homogenised at all radii, showing similar surface density profiles.
The results from TNG100 illustrated in this paper and in previous works in literature \citep[e.g.,][]{Pillepich2018MNRAS,Tacchella2019MNRAS,Pulsoni2021A&A} support the possible scenario in which massive systems assembled across their cosmic histories mainly via major mergers \citep[see also][]{SonnenfeldNipotiTreu2017MNRAS}, in contrast with some previous theoretical \citep[e.g.,][]{Naab2009ApJl,Oser2010ApJ,Hilz2013MNRAS} and observational \citep[e.g.,][]{Bell2006ApJ,McIntosh2008MNRAS} studies which, instead, endorse a minor merger-driven evolution for ETGs.
\end{itemize}


For the future, we plan to extend the analysis to other physical properties, like chemical abundances of individual elements. In order to provide a more complete scenario behind the cosmic evolution of the ETGs that we observe in the present-day Universe, in simulations we will study the merger history of individual galaxies,  considering the evolution of the spatial distribution of their stellar properties. Finally, we will make use of the newest TNG50 simulation of IllustrisTNG which, though characterised by a smaller physical volume and thus a lower statistics, benefits from a higher mass resolution that could allow us to make a more reliable comparison at smaller scales of galaxies with data from current and upcoming surveys.

\section*{Acknowledgements}
We are grateful to M. Auger and M. Bernardi for helpful pieces of advice and comments, which remarkably enriched the analysis presented in this work.
We sincerely thank the referee A. Pillepich for her feedback and suggestions that considerably contributed to improve the quality of this manuscript.
CC acknowledges useful discussions with F. Ardila, C. Bacchini, F. Belfiore, C. D'Eugenio, G. Iorio, M. Mingozzi, R. Pascale, S. Quai, and S. Tacchella. This material is based upon work supported by the National Science Foundation under Grant No. 1714610.

\section*{Data availability}
The data underlying this article regarding the selection of galaxy samples and the radial profiles will be shared upon request to the corresponding author.
IllustrisTNG data are publicly available at \url{https: //www.illustris-project.org/data/} \citep{NelsonMNRAS2021}.
The MaNGA galaxy sample is drawn from an extended version of the NASA-Sloan Atlas \citep[NSA {\ttfamily v1\_0\_1}, \url{https://www.sdss.org/dr15/manga/manga-target-selection/nsa/};][]{Blanton2011AJ} catalogue. MaNGA DR15 data are taken from \url{https://www.sdss.org/dr15/manga/manga-data/} \citep[][]{Aguado2019ApJS}. The stellar masses for MaNGA galaxies are included in the \textsc{UPenn\_PhotDec\_MsSTAR} catalogue \citep[\url{http://alan-meert-website-aws.s3-website-us-east-1.amazonaws.com/fit_catalog/download/index.html;}][]{Meert2015MNRAS}.
To check the impact of the adopted selection for the MaNGA sample, we relied on the morphological type  presented in the  \textsc{MaNGA Morphology Deep Learning DR15 catalog} \citep[\url{https://www.sdss.org/dr15/data_access/value-added-catalogs/?vac_id=manga-morphology-deep-learning-dr15-catalogue};][]{Fischer2019MNRAS}.



\bibliographystyle{mnras}
\bibliography{bibliography} 




\appendix

\newpage
\onecolumn
\section{Computing the angular diameter distances to convolve TNG100 maps}
\label{app:angular_distances}

As described in \autoref{ssec:radial_profiles_tng}, in order to account for the effects of the MaNGA resolution on simulated galaxies, for each stellar property we consider two limit-case profiles: for a given simulated ETG, one profile is computed directly from the original $2\mathrm{D}$ stellar property map, i.e. the \emph{unconvolved profile}, while the other profile is derived from a map previously convolved with a $2\mathrm{D}$ Gaussian filter kernel $\sigma_\mathrm{kernel}$, i.e. the \emph{convolved profile}.
To compute the kernel of each simulated ETG, we use  \autoref{eq:kernel}, where $\mathcal{R}_\mathrm{TNG}=1\,\kpc$ is the TNG100 resolution of the original maps, while $\mathcal{R}_{\mathrm{MaNGA},i}=\sin(\mathrm{PSF}_\mathrm{MaNGA})d_{A,i}$ is the resolution that the $i$-th TNG100 ETG map should have if it were observed as a MaNGA galaxy, and depends on $\mathrm{PSF}_\mathrm{MaNGA}=2.5^\dprime$ and the angular diameter distance $d_{A,i}$ of the $i$-th TNG100 galaxy.
To measure  $d_{A,i}$ we rely on the {\ttfamily angular\_diameter\_distance} function of the Python package \textsc{Astropy}, that takes in input the redshift of the $i$-th source.

\begin{figure}
    \centering
    \includegraphics[width=.4\columnwidth]{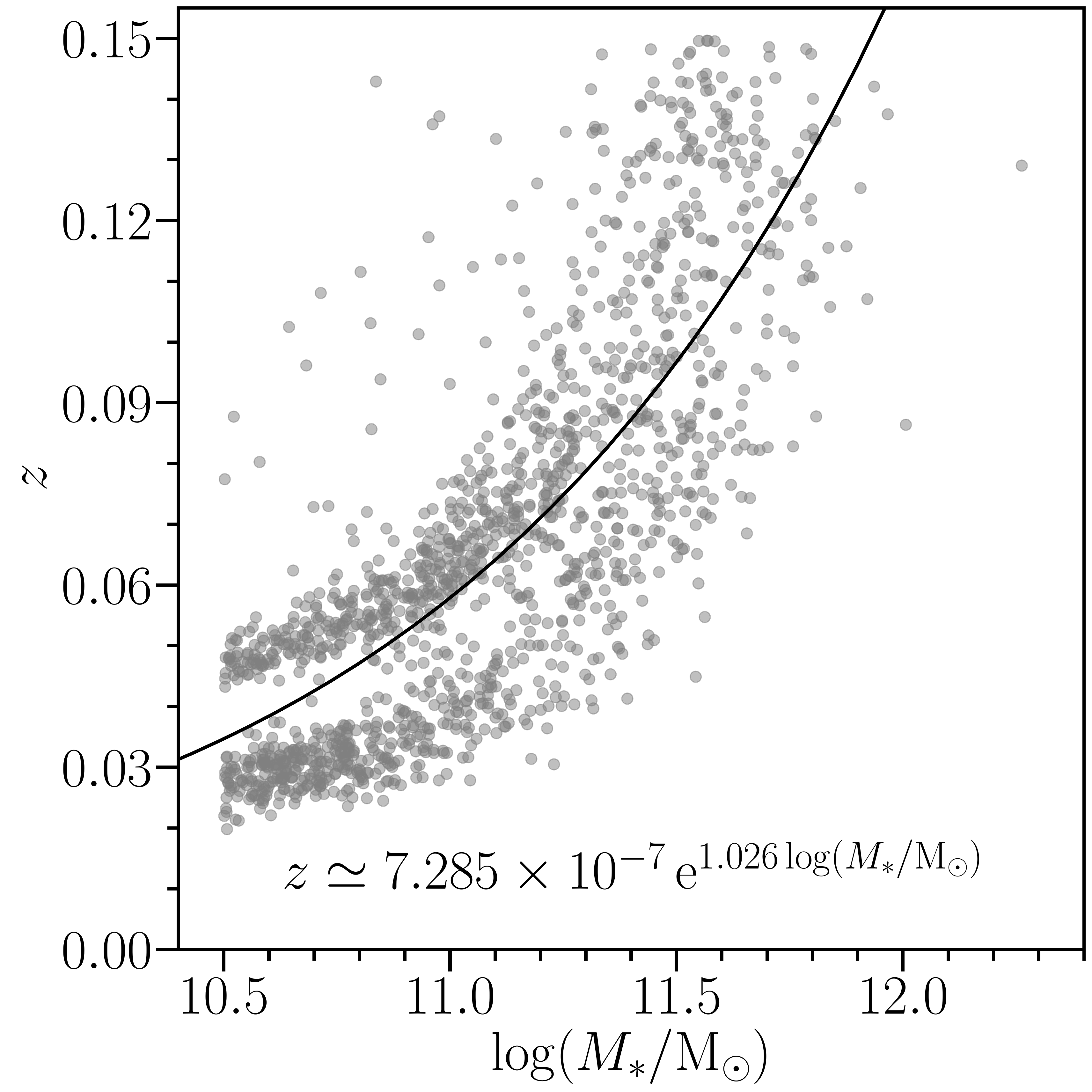}
    \caption{The redshift${-}$stellar mass  distributions of the MaNGA ETG sample. The black solid curve traces the fit of the distribution. The corresponding fit functional form is reported. The two stripes of dots trace the Primary sample (the lower cloud) and the Secondary sample (the upper cloud) of the MaNGA sample.}
    \label{fig:z_mass_planes}
\end{figure}

Since MaNGA was built in such a way that the most massive galaxies are located at higher redshifts, we fit the $z{-}\mstar$ distribution of the MaNGA sample considered and we assign to each simulated galaxy the corresponding redshift value depending on its stellar mass.
The functional form adopted for fitting the $z{-}\mstar$ distributions in MaNGA is

\begin{equation}
\centering
z=a\, e^{b\log(\mstar/\msun)},
\end{equation}
where $a$ and $b$ are the two parameters used for the fit. In \autoref{fig:z_mass_planes}, the $z{-}\mstar$ scatter distributions with the corresponding fit are shown.
Thus, the fit is used to assign redshifts to the simulated ETGs to compute their angular diameter distance, and then the kernel used to convolve their stellar property maps.
We remind the reader that simulated galaxies are extracted from the $z=0.1$ snapshot of TNG100-1. However, to make the comparison with MaNGA as fair as possible, for our scope, we ignore this information for convolving the maps and we reassign to each simulated ETG a new redshift corresponding to its stellar mass according to the fit of the MaNGA distribution on the $z{-}\mstar$ plane.

\section{Comparing profiles in number-density-based stellar mass bins}
\label{app:number_density_cases}

\begin{figure}
    \centering
    \includegraphics[width=0.5\columnwidth]{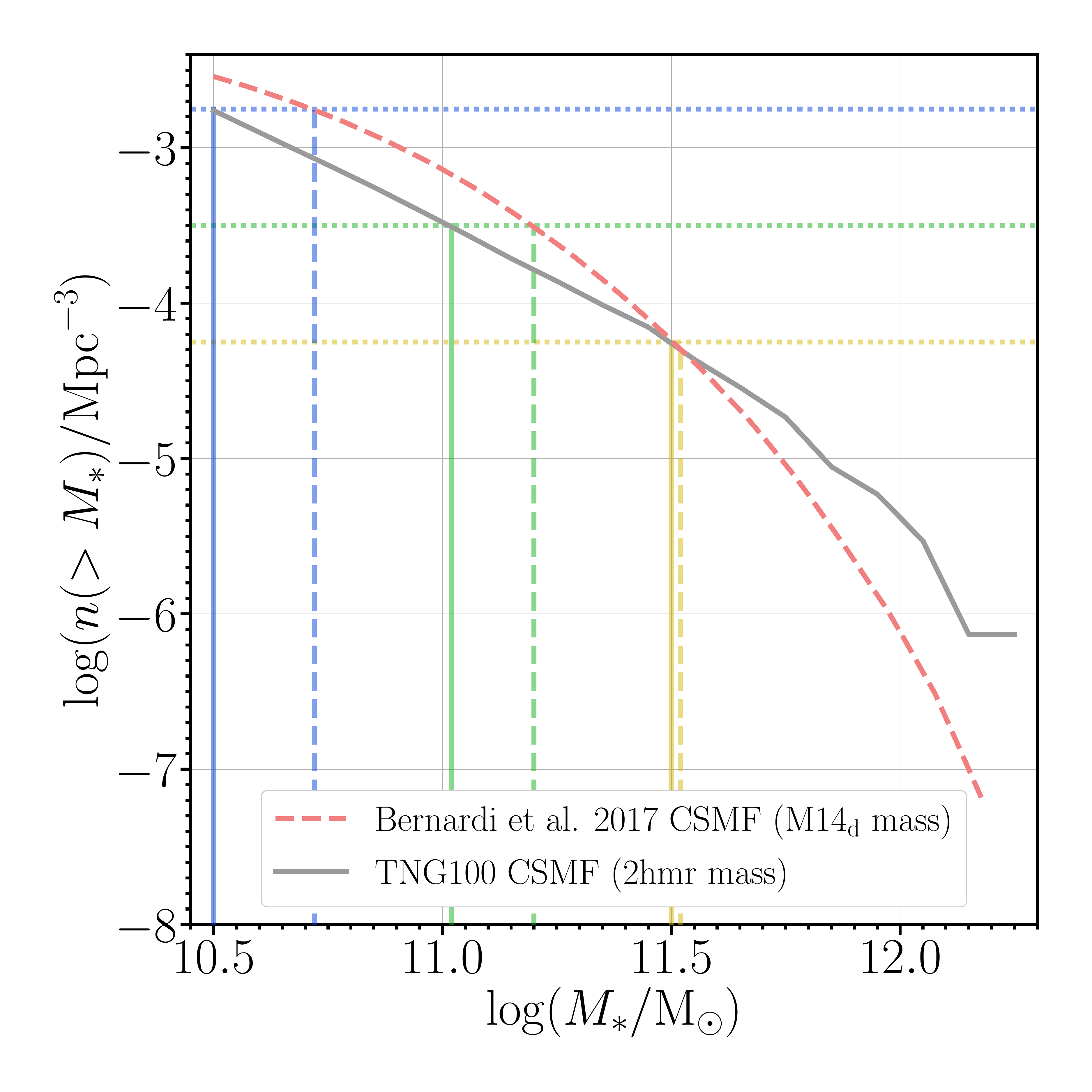}
    \caption{CSMFs for the TNG100 (grey solid curve) and the MaNGA (red dashed curve) samples. Horizontal dotted lines indicate out number-density-based bins:
    $\log (n/\mathrm{Mpc}^{-3})=-2.75$ (blue),
    $\log (n/\mathrm{Mpc}^{-3})=-3.50$ (green),
    $\log (n/\mathrm{Mpc}^{-3})=-4.75$ (yellow). Vertical lines indicate the corresponding stellar mass values for TNG100 (solid lines) and for MaNGA (dashed lines).}
    \label{fig:cmfs_bins}
\end{figure}

In \autoref{sec:manga_tng_results} we presented the radial profiles of MaNGA and TNG100 ETG stellar properties in the three stellar mass bins $10.5\leq\log(\mstar/\msun)<11$, $11\leq\log(\mstar/\msun)<11.5$, and $\log(\mstar/\msun)\geq11.5$.
We repeated here the same analysis building stellar mass bins at \emph{fixed number density}. Specifically, we compute the stellar mass function (SMF) for our TNG100 galaxy sample, whereas for MaNGA we use Table 1 of \citet{Bernardi2017MNRAS} and adopt the observed (i.e. error-broadened) SMF associated with the Dusty ($\boldsymbol{\Phi}_\mathrm{Obs}^\mathrm{M14_{d}}$) mass estimates from \citet{Mendel2014ApJS} with the SerExp photometry of \citet{Meert2015MNRAS}. These are the same stellar masses as used for our MaNGA ETGs. Thus, we compute the cumulative stellar mass functions (CSMFs) for both MaNGA and TNG100 as the sum of the number counts of galaxies with stellar masses greater than a given value $M_{*,i}$:
\begin{equation}
n(> M_{*,i})=\int_{ M_{*,i}}^{+\infty}\boldsymbol{\Phi}( M_*^\prime)\,\mathrm{d} M_*^\prime.
\end{equation}
\autoref{fig:cmfs_bins} shows the CSMFs for TNG100 and MaNGA samples. We adopt three bins in number density as listed in \autoref{tab:manga_illustristng_cmfs}. Number density bins could be more robust against possible mismatches in the stellar mass measurements between observations and simulations. \autoref{tab:manga_illustristng_cmfs} indicates the values of the stellar masses in both TNG100 and MaNGA samples that correspond to our three number density bins:
$-3.50<\log (n/\mathrm{Mpc}^{-3})\leq-2.75$,
$-4.25<\log (n/\mathrm{Mpc}^{-3})\leq-3.50$,
and $\log (n/\mathrm{Mpc}^{-3})\leq-4.25$.
The bounds of the number density bins, listed in \autoref{tab:manga_illustristng_cmfs}, are such that the corresponding stellar mass values for simulated ETGs, i.e. $\log(\mstar/\msun)=10.5, 11.02, 11.5$, almost coincide with the values of stellar masses used in \autoref{sec:manga_tng_results}.

\begin{table}
\centering
\caption{Stellar mass values corresponding to the bounds of the number density bins. Column 1: number density. Column 2: stellar mass for the MaNGA sample.
Column 3: stellar mass for the TNG100 sample. Stellar masses are in units of $\msun$.}
\begin{tabular}{ccc}
\toprule
\toprule
\addlinespace
$\log (n/\mathrm{Mpc}^{-3})$ & $\log M_*,_\mathrm{MaNGA}$ &
$\log M_*,_\mathrm{TNG}$ \\
\addlinespace
\midrule
\addlinespace
$-2.75$ & 10.72 & 10.50 \\
\addlinespace
$-3.50$ & 11.20 & 11.02 \\
\addlinespace
$-4.25$ & 11.52 & 11.50 \\
\addlinespace
\bottomrule
\bottomrule
\end{tabular}
\label{tab:manga_illustristng_cmfs}
\end{table}

\begin{figure*}
    \centering
    \includegraphics[width=.95\textwidth]{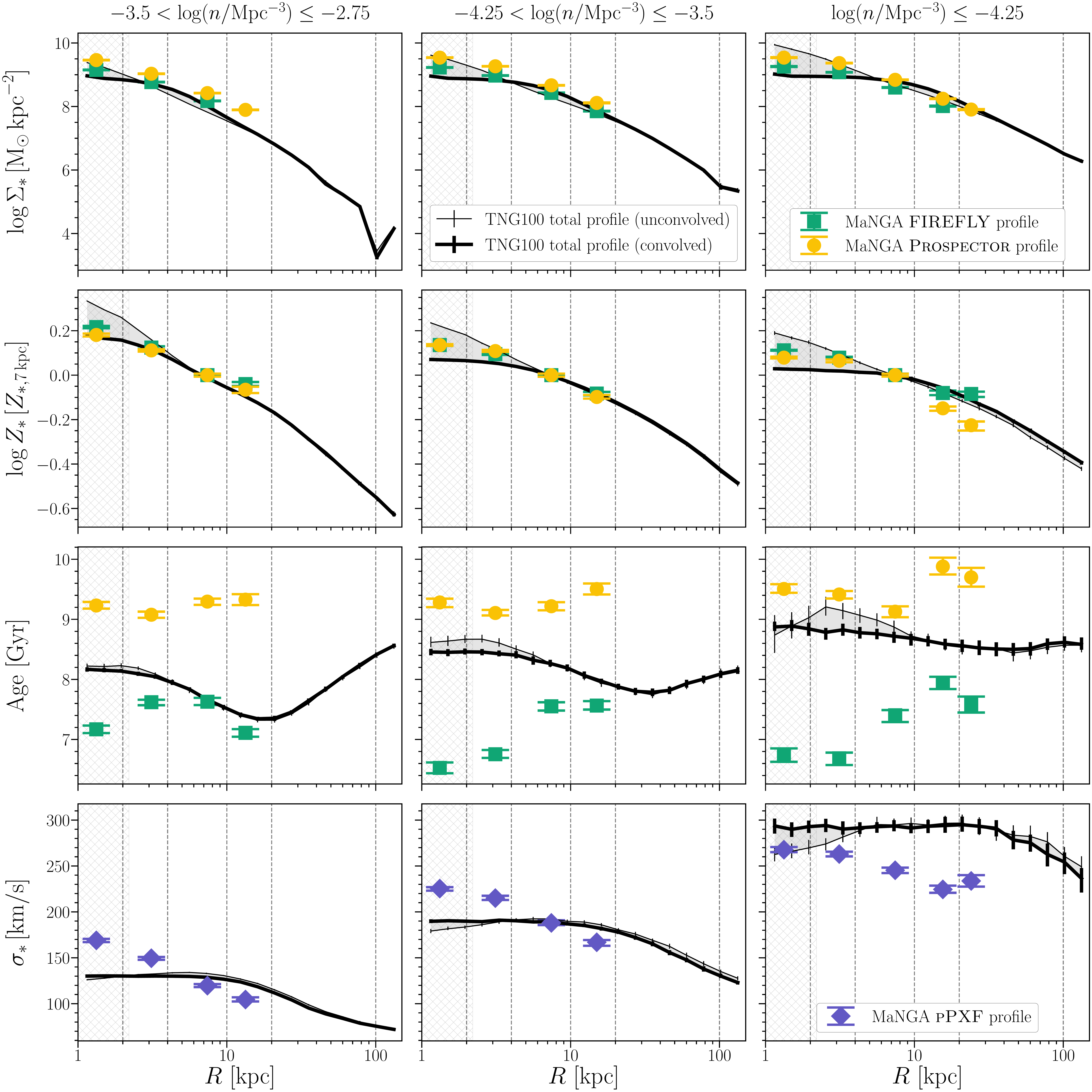}
    \caption{Same as \autoref{fig:profiles_mass_dusty}, but in number-density-based stellar mass bins.}
    \label{fig:profiles_number_density_dusty}
\end{figure*}

As illustrated in \autoref{fig:profiles_number_density_dusty}, the most evident exceptions of adopting the number-density-based stellar mass bins concern the first two bins of the stellar properties analysed. In particular, the discrepancy between the radial profiles of the stellar mass surface density for the observed and simulated ETGs increases, highlighting a tendency of the MaNGA sources to assume slightly higher values, with respect to those shown in \autoref{fig:profiles_mass_dusty}.
A similar behaviour, but less significant, is also found for the stellar metallicity and age distributions.
Instead, concerning velocity dispersion, in the first two bins the MaNGA profiles are systematically shifted up by a factor of around $20{-}30\,\kms$
All these discrepancies are caused by the fact that, by removing from the MaNGA sample ETGs with $\logm\leq10.78$ in the first bin, and considering  galaxies with $\log(\mstar/\msun)>11.2$ in the second bin, the median profiles of the observed sources tend to assume higher values than the counterparts presented in \autoref{fig:profiles_mass_dusty}.

Since we expect number density bins to more closely approximate a halo-mass-based comparison, the aforementioned differences could be an indication of a mismatch between MaNGA and TNG100 in the stellar-to-halo mass relation.
However, qualitatively speaking, the selection in number density bins does not affect remarkably the overall fashion of the profiles for all the stellar properties, showing similar radial distributions as those illustrated in \autoref{fig:profiles_mass_dusty_inex}.

\begin{figure*}
    \centering
    \includegraphics[width=.95\textwidth]{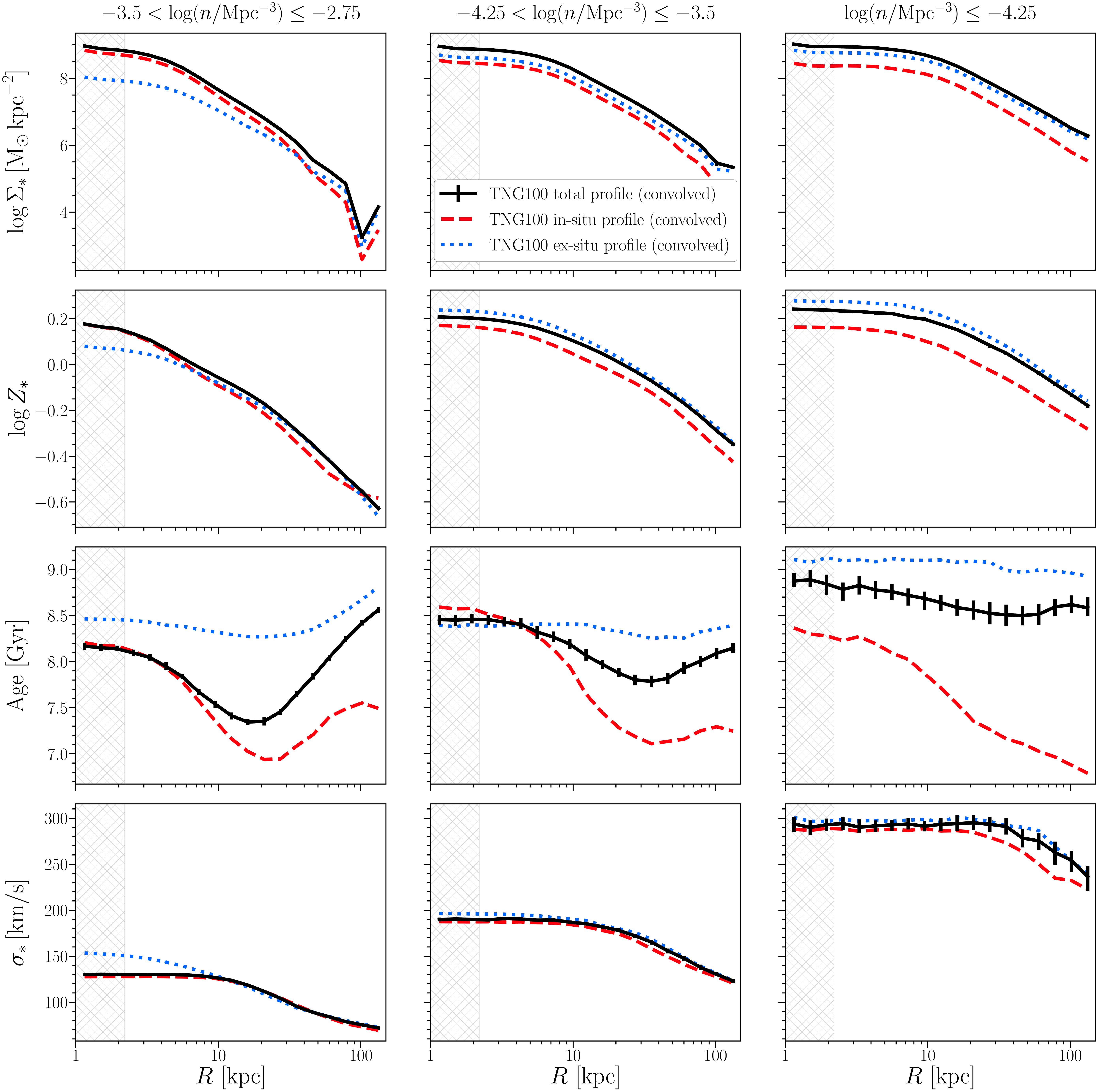}
    \caption{Same as \autoref{fig:profiles_mass_dusty_inex}, but in number-density-based stellar mass bins.}
    \label{fig:profiles_number_density_dusty_inex}
\end{figure*}

\bsp	
\label{lastpage}
\end{document}